\documentclass{amsart}
\usepackage{amssymb}
\usepackage{amsmath}
\usepackage{amsfonts}

\theoremstyle{plain}

\numberwithin{equation}{section}
\input{tcilatex}

\begin{document}
\title[A METHOD FOR FINDING ZEROS OF SET-VALUED OPERATORS]{A
PROXIMAL-PROJECTION METHOD FOR FINDING ZEROS OF SET-VALUED OPERATORS}
\author[Dan Butnariu]{Dan Butnariu$^{\ast }$}
\address{Department of Mathematics\\
University of Haifa\\
31905 Haifa, Israel}
\email{dbutnaru@math.haifa.ac.il}
\author[G\'{a}bor Kassay]{G\'{a}bor Kassay$^{\ast \ast }$}
\address{Faculty of Mathematics \\
Babes-Bolyai University\\
3400 Cluj-Napoca, Romania}
\email{\textit{\ }kassay@math.ubbcluj.ro}
\date{February 4, 2007}
\keywords{Bregman distance, Bregman projection, $D_{f}$-antiresolvent, $%
D_{f} $-nonexpansivity pole, $D_{f}$-coercive operator, $D_{f}$-firm
operator, $D_{f}$-nonexpansive operator, $D_{f}$-resolvent, firmly
nonexpansive operator, Legendre function, maximal monotone operator,
monotone operator, nonexpansive operator, proximal mapping, proximal point
method, proximal projection method, relative projection, sequentially
consistent function, projected subgradient method, Tikhonov-Browder
regularization, strongly monotone operator, uniformly convex function,
variational inequality.\smallskip \\
$^{\ast }$Dan Butnariu's work on this paper was partially done during his
visit within The Graduate Center of the City University of New York and he
gratefully acknowledges Professor Gabor Herman's support and the informative
discussions and activities in which he was invited to take part.\smallskip \\
$^{\ast \ast }$Gabor Kassay gratefully acknowledges the support of Cesarea
Benjamin de Rothschild Foundation during his May 2006 visit within the
Department of Mathematics of the University of Haifa, Israel, when part of
this research was completed.}

\begin{abstract}
In this paper we study the convergence of an iterative algorithm for finding
zeros with constraints for not necessarily monotone set-valued operators in
a reflexive Banach space. This algorithm, which we call the
proximal-projection method is, essentially, a fixed point procedure and our
convergence results are based on new generalizations of Lemma Opial. We show
how the proximal-projection method can be applied for solving ill-posed
variational inequalities and convex optimization problems with data given or
computable by approximations only. The convergence properties of the
proximal-projection method we establish also allow us to prove that the
proximal point method (with Bregman distances), whose convergence was known
to happen for maximal monotone operators, still converges when the operator
involved in it is monotone with sequentially weakly closed graph.
\end{abstract}

\maketitle

\section{Introduction}

In that follows $X$ denotes a real reflexive Banach space with norm $%
\left\Vert \cdot \right\Vert $ and $X^{\ast }$ denotes the (topological)
dual of $X$ with the dual norm $\left\Vert \cdot \right\Vert _{\ast }$. Let $%
f:X\rightarrow \mathbb{(-\infty },+\infty ]$ be a proper, lower
semicontinuous convex function with domain $\mathrm{dom}\,f$. Then, the
Fenchel conjugate $f^{\ast }:X^{\ast }\rightarrow \mathbb{(-\infty },+\infty
]$ is also a proper lower semicontinuous convex function and $f^{\ast \ast
}:=(f^{\ast })^{\ast }=f$. We assume that $f$ is a Legendre function in the
sense given to this term in \cite[Definition 5.2]{BauBorCom-Essential}, that
is, $f$ is essentially smooth and essentially strictly convex. Then,
according to \cite[Theorem 5.4]{BauBorCom-Essential}, the function $f^{\ast
} $ is a Legendre function too. Moreover, by \cite[Theorem 5.6 and Theorem
5.10]{BauBorCom-Essential}, both functions $f$ and $f^{\ast }$ have domains
with nonempty interior, are (G\^{a}teaux) differentiable on the interiors of
their respective domains,%
\begin{equation}
\mathrm{ran}\,\nabla f=\mathrm{dom}\,\nabla f^{\ast }=\mathrm{int\,dom}%
\,f^{\ast }=\mathrm{dom}\,\partial f^{\ast },  \label{domf'}
\end{equation}%
\begin{equation}
\mathrm{ran}\,\nabla f^{\ast }=\mathrm{dom}\,\nabla f=\mathrm{int\,dom}\,f=%
\mathrm{dom}\,\partial f,  \label{domf*'}
\end{equation}%
and%
\begin{equation}
\nabla f=(\nabla f^{\ast })^{-1}.  \label{f'f*}
\end{equation}

With the function $f$ we associate the function $W_{f}:X^{\ast }\times
X\rightarrow (-\infty ,+\infty ]$ defined by%
\begin{equation}
W_{f}(\xi ,x)=f(x)-\left\langle \xi ,x\right\rangle +f^{\ast }(\xi ).
\label{Wf}
\end{equation}%
By the Young-Fenchel inequality the function $W_{f}$ is nonnegative and $%
\mathrm{dom}\,W_{f}=\left( \mathrm{dom}\,f^{\ast }\right) \times \left( 
\mathrm{dom}\,f\right) .$ It is known (see \cite{Alber1993}, \cite%
{alberGenproj}, \cite{BauBor-LegendreFcts}, \cite{ButResSurvey} and see also
Section 2 below) that, for any nonempty closed convex set $E$ contained in $%
X\ $such that $E\,\cap \,\mathrm{int}\,\mathrm{dom}\,f\neq \varnothing ,$
the function ${\mathrm{Proj}}_{E}^{f}:\mathrm{int\,dom}\,f^{\ast
}\rightarrow X$ given by%
\begin{equation}
\mathrm{Proj}_{E}^{f}\xi =\arg \min \left\{ W_{f}(\xi ,x):x\in E\right\} ,
\label{proxox}
\end{equation}%
is well defined and its range is contained in $E\cap \mathrm{int}\,\mathrm{%
dom}\,f.$ In fact, this function is a particular proximal projection in the
sense given to this term in \cite{BauBorCom-BregMon} which was termed in 
\cite{ButResSurvey} projection onto $E$\ relative to $f$ because in the
particular case when $X$ is a Hilbert space and $f(x)=\textstyle%
\frac{1}{2}%
\left\Vert x\right\Vert ^{2}$ the vector $\mathrm{Proj}_{E}^{f}\xi $
coincides with the usual (metric) projection of $\xi $ onto $E.$

In this paper we are interested in the following problem:\smallskip

\textbf{Problem 1.1: }\textit{Given an operator }$A:X\rightarrow 2^{X^{\ast
}}$\textit{\ and a nonempty closed subset }$C$\textit{\ of }$X$\textit{\
such that}%
\begin{equation}
\varnothing \neq C\cap \mathrm{dom}\,A\subseteq \mathrm{int}\,\mathrm{dom}%
\,f,  \label{consist1}
\end{equation}%
\textit{find }$x\in C$\textit{\ such that }$0^{\ast }\in Ax,$\textit{\ where 
}$0^{\ast }$\textit{\ denotes the null vector in }$X^{\ast }.\smallskip $

\noindent Our purpose is to discover sufficient conditions for the following
iterative procedure, which we call \textit{the\ proximal-projection method},%
\begin{eqnarray}
x^{0} &\in &C_{0}\cap \mathrm{dom}\,A\cap \mathrm{int}\,\mathrm{dom}\,f\text{%
\quad and}  \label{alg} \\
x^{k+1} &\in &\mathrm{Proj}_{C_{k+1}\cap \mathrm{dom}\,A}^{f}(\nabla
f(x^{k})-Ax^{k}),\text{ }\forall k\in \mathbb{N}\text{,}  \notag
\end{eqnarray}%
to generate approximations of solutions to the Problem 1.1 when $\left\{
C_{k}\right\} _{k\in \mathbb{N}}$ is a sequence of closed convex subsets of $%
X$ approximating weakly (see Definition 4.2 below) the set $C$ under the
following conditions of compatibility of $f$ and $C_{k}$ with the data of
Problem 1.1:\smallskip

\textbf{Assumption 1.1: }\textit{For each}\textbf{\ }$k\in \mathbb{N}$%
\textit{, the set} $C_{k}\cap \mathrm{dom}\,A$ \textit{is convex and closed,}%
\begin{equation}
C\subseteq C_{k}\text{\quad and\quad }(\nabla f-A)(C_{k})\subseteq \mathrm{%
int}\,\mathrm{dom}\,f^{\ast }.  \label{consistency}
\end{equation}

\noindent In this paper (\ref{consist1}) and Assumption 1.1 are standing
assumptions, even if not explicitly mentioned, whenever we refer to the
Problem 1.1 or to the proximal-projection method. In view of (\ref{consist1}%
) and (\ref{consistency}) the sets $C_{k}\cap \mathrm{dom}\,A\cap \mathrm{int%
}\,\mathrm{dom}\,f$ are necessarily nonempty and, consequently, $(\nabla
f-A)(C_{k})$ is nonempty too. This fact, Assumption 1.1 and Lemma 2.1 below
which ensures that%
\begin{equation*}
\mathrm{dom}\,\mathrm{Proj}_{C_{k}\cap \mathrm{dom}\,A}^{f}=\mathrm{int}\,%
\mathrm{dom}\,f^{\ast }\text{ and }\mathrm{ran}\,\mathrm{Proj}_{C_{k}\cap 
\mathrm{dom}\,A}^{f}\subseteq C_{k}\cap \mathrm{dom}\,A\cap \mathrm{int}\,%
\mathrm{dom}\,f,
\end{equation*}%
taken together, guarantee that the procedure of generating sequences in (\ref%
{alg}) is well defined.

The proximal-projection method described above is a natural generalization
of the method of finding zeros of linear operators due to Landweber \cite%
{landweber}, of Shor's \cite{shor1962} (see also \cite{shorbook}) and
Ermoliev's \cite{ermoliev} "gradient descent" methods for finding
unconstrained minima of convex functions and of the "projected-subgradient"
method for finding constrained minima of convex functions studied by Polyak 
\cite{polyak}. These methods inspired the construction of a plethora of
algorithms for finding zeros of various operators as well as for other
purposes. Among them are the algorithms presented in \cite{Alber1993}, \cite%
{alber3}, \cite{alberStab}, \cite{AlbButKas}, \cite{AlbButRya2001}, \cite%
{AlbButRya2004}, \cite{AlbButRya2005}, \cite{AlberGuerre2001}, \cite%
{alber-ius-sol}, \cite{alb-kar-lit}, \cite{alb-nash}, \cite{alber-rei}, \cite%
{bak/polyak}, \cite{BauBorCom-BregMon}, \cite{bruck1}, \cite{ButResOptim}, 
\cite{ButResSurvey}, \cite{ruszczynski} which, in turn, inspired this
research. The main formal differences between the proximal-projection method
(\ref{alg}) and its already classical counterparts developed in the 50-ies
and 60-ies consist of the use of the proximal projections instead of metric
projections and of projecting not on the set $C$ involved in Problem 1.1,
but on some convex approximations $C_{k}$ of it. The use of proximal
projections instead of metric projections is mostly due to the fact that
metric projections in Banach spaces which are not Hilbertian do not have
many of those properties (like single valuedness and nonexpansivity) which
make them so useful in a Hilbert space setting for establishing convergence
of algorithms based on them. As far as we know, the idea of using proximal
projections instead of metric projections in projected-subgradient type
algorithms goes back to Alber's works \cite{Alber1993}, \cite{alberGenproj}.

As we make clear in Section 3.3, there is an intimate connection between the
proximal-projection method and the well-known proximal point method with
Bregman distances -- see (\ref{proxpt}). The proximal point method with
Bregman distances considered in this paper is itself a generalization of the
classical proximal point algorithm developed since the ninety-fifties by
Krasnoselskii \cite{krasnoselskii}, Moreau \cite{Moreau1962}, \cite%
{Moreau1963}, \cite{Moreau1965}, Yosida \cite{yoshida}, Martinet \cite%
{martinet}, \cite{martinet-1} and Rockafellar \cite{RocProxPt}, \cite%
{RocAugLag} among others (see \cite{lemaire} for a survey of the literature
concerning the classical proximal point method). It emerged from the works
of Erlander \cite{erlander}, Eriksson \cite{eriksson}, Eggermont \cite%
{eggermont} and Eckstein \cite{eckstein} who studied various instances of
the algorithm in $\mathbb{R}^{n}.$ Its convergence analysis in Banach spaces
which are not necessarily Hilbertian was initiated in \cite{BurSch}, \cite%
{ButIusProxPt} and \cite{kassay} (see \cite{ButIus-Book} and \cite{BurSch}
for related references on this topic). Lemma 3.5 shows that, in our setting,
the proximal point method with Bregman distances is a particular instance of
the proximal-projection method.

Computing projections, metric or proximal, onto a closed convex set with
complicated geometry is, in itself, a challenging problem. It requires (see (%
\ref{proxox})) solving convex nonlinear programming problems with convex
constraints. Specific techniques for finding proximal projections are
presented in \cite{AlbBut}, \cite{BauCom-BregBestAp} and \cite%
{butnariu-ius-res}. It is obvious from these works that it is much easier to
find proximal projections onto sets with simple geometry like, for instance,
hyperplanes, half spaces or finite intersections of such sets. These facts
naturally led to the question of whether it is possible to replace in the
process of computing iterates of metric or proximal projections algorithms
the constraint set $C$ by some approximations $C_{k}$ of it whose geometry
is simple enough to allow relatively easy calculation of the required metric
or proximal projections at each iterative step $k$. That this approach is
sound is quite clear from the works of Mosco \cite{mosco} and its subsequent
developments due to Attouch \cite{attouch}, Aubin and Frankowska \cite%
{AubFrank}, Dontchev and Zolezzi \cite{DonZol}, and from the studies of
Liskovets \cite{liskovets3}, \cite{liskovets4}. Its main difficulty in the
case of the proximal-projection method is that the approximations $C_{k}$
one uses should converge to $C$ in a manner that ensures stable convergence
of the algorithm to solutions of the problem. For some variants of the
proximal-projection method, types of convergence of the sets $C_{k}$ to $C$
which are sufficiently good for this purpose are presented in \cite%
{alberStab}, \cite{AlbButRya2001}, \cite{AlbButRya2005}, \cite{alb-kar-lit}, 
\cite{alb-nash}. They mostly are relaxed forms of Hausdorff metric
convergence. It was shown in \cite{AlbButRya2004} that, in some
circumstances, fast Mosco convergence of the sets $C_{k}$ to $C$ (see \cite[%
Definition 2.1]{AlbButRya2004}), a form of convergence significantly less
demanding than Hausdorff convergence, is sufficient to make the
proximal-projection method (\ref{alg}) applied to variational inequalities
converge. As we show below, these convergence requirements in the case of
the proximal-projection method (\ref{alg}) can be further weakened. In fact,
in our convergence theorems for the proximal-projection method we only
require weak Mosco convergence of the sets $C_{k}$ to $C$ (see Definition
4.3) and this is significantly demanding than Hausdorff metric or fast Mosco
convergence.

The purpose of this work is to find general conditions which guarantee that
the proximal-projection method converges weakly or strongly to solutions of
the Problem 1.1. Observe that there is no apparent connection between the
data of Problem 1.1 and the function $f$ involved in the definition of the
proximal-projection method. Our main question is how the function $f$ should
be chosen in order to ensure (weak or strong) convergence of the
proximal-projection method to solutions of Problem 1.1 without excessively
conditioning the problem data. The function $f$ is a parameter of the
proximal-projection method whose appropriate choice, as we show below, can
make the procedure converge to solutions of Problem 1.1 even if the problem
data are quite "bad" in the sense that they do not have some, usually
difficult to verify in practice properties like maximal monotonicity, strict
monotonicity, various forms of nonexpansivity, continuity or closedness
properties of some kind or another. Theorem 4.1 and Theorem 5.1 are our
responses to the question posed above.

Theorem 4.1 shows that for guaranteeing that the proximal-projection method
produces weak approximations of solutions for Problem 1.1 it is sufficient
to chose a function $f$ which, besides the conditions (\ref{consist1}) and (%
\ref{consistency}) which are meant to make the procedure consistent with the
problem data, should satisfy some requirements which, most of them, are
common features of the powers of the norm $\left\Vert \cdot \right\Vert ^{p}$
with $p>1$ in uniformly convex and smooth Banach spaces. The only somehow
outstanding condition which we require for $f$ is that it should be such
that the operator $A$ involved in the problem be $D_{f}$-coercive on $C$ or,
if the set $C$ is approximated by sets $C_{k},$ then $D_{f}$-coercivity of $%
A $ should happen on the union of those sets. $D_{f}$-coercivity, a notion
introduced in this paper (see Definition 3.2), is a generalization of the
notion of firm nonexpansivity for operators in a Hilbert space (see (\ref%
{fnonexp})). Although in Hilbert spaces provided with the function $f=%
\textstyle%
\frac{1}{2}%
\left\Vert \cdot \right\Vert ^{2}$ this notion coincides with the notion of
firm nonexpansivity and, also, with the notion of $D_{f}$-firmness
introduced in \cite{BauBorCom-BregMon} (see Definition 3.3 below), outside
this particular setting the notions of $D_{f}$-coercivity and $D_{f}$%
-firmness complement each other (cf. Section 3.2). In Section 4.3 we present
several corollaries of Theorem 4.1 and examples which clearly show that
fitting a function $f$ to the specific data of Problem 1.1 may came
naturally in many situations. If $X$ is a Hilbert space and $A$ is a firmly
nonexpansive operator, then the natural choice is $f=\textstyle%
\frac{1}{2}%
\left\Vert \cdot \right\Vert ^{2}$ (see Corollary 4.1). In this case the
convergence of the proximal-projection method happens to be strong if the
set of solutions of the problem has nonempty interior. If in Problem\ 1.1 we
have $C=X,$ then the problem is equivalent to that of finding a zero for the
operator $\nabla f-\nabla f\circ A_{f},$ where $A_{f}$ is the $D_{f}$%
-resolvent of $A$ (a notion introduced in \cite{BauBorCom-BregMon} - see
also (\ref{res})) and application of the proximal-projection method with $%
C_{k}=X$ to $\nabla f-\nabla f\circ A_{f}$ is no more and no less than the
proximal point method with Bregman distances mentioned above. The operator $%
\nabla f-\nabla f\circ A_{f}$ is $D_{f}$-coercive whenever the operator $A$
is monotone (cf. Lemma 3.4). This leads us to the application of Theorem 4.1
to the operator $\nabla f-\nabla f\circ A_{f}$ which is Corollary 4.2. It
shows that the proximal point algorithm with Bregman distances converges
subsequentially weakly (and when $A$ has a single zero, sequentially weakly)
for a large class of functions $f,$ whenever $A$ is monotone and provided
that its graph is sequentially weakly closed (as happens, for instance, when 
$\func{Graph}A$ is convex and closed in $X\times X^{\ast }$). It seems to us
that this is the first time when weak convergence of the proximal point
algorithm with Bregman distances is proved without requiring maximal
monotonicity of $A.$ The proximal-projection method is also a tool for
solving monotone variational inequalities via their Tikhonov-Browder
regularization. This is shown by Corollary 4.3, another consequence of
Theorem 4.1. Corollary 4.3 also asks for the monotone operator $%
B:X\rightarrow 2^{X^{\ast }}$ involved in the variational inequality to be
such that $\nabla f-\nabla f\circ \mathrm{Proj}_{C}^{f}\circ \left[
(1-\alpha )\nabla f-B\right] $ is $D_{f}$-coercive (or, equivalently, such
that $\mathrm{Proj}_{C}^{f}\circ \left[ (1-\alpha )\nabla f-B\right] $ to be 
$D_{f}$-firm). This happens in many situations of practical interest.
Several such situations are described in the Examples 4.1 and 4.2.

A careful analysis of the proof of Theorem 4.1 reveals the fact that the
proximal-projection method (\ref{alg}) is a procedure of approximating fixed
points for the operator $\mathrm{Proj}_{C}^{f}\circ \left( \nabla f-A\right) 
$ by iterating the operator. A customary tool of proving convergence of such
algorithms in Hilbert spaces is the already classical Opial Lemma \cite[%
Lemma 2]{opial}. Unfortunately, this result can not be extrapolated into a
nonhilbertian setting in its original form. Our Theorem 4.1 is based on
Proposition 4.1, a generalization of the Opial Lemma which works in
reflexive Banach spaces and which is of interest by itself. If the Banach
space $X$ has finite dimension, Proposition 4.1 can be substantially
improved -- see Proposition 5.1. Thus, in spaces with finite dimension we
can also improve Theorem 4.1 by dropping some of the requirements made on
the problem data. This is, in fact, our Theorem 5.1 which guarantees
convergence of the proximal-projection method with less demanding conditions
than closedness of the graph of $A.$ Accordingly, in finite dimensional
spaces the conclusions of Corollaries 4.2 and 4.3 can be reached at lesser
cost for the operators involved in them as shown by Corollaries 5.1 and 5.2,
respectively.

This paper continues and develops a series of concepts, methods and
techniques initiated in \cite{Alber1993}, \cite{BauBor-LegendreFcts}, \cite%
{BauBorCom-BregMon}, \cite{ButIus-Book}, \cite{ButResSurvey} and \cite%
{kassay}. In Sections 2 and 3 we present in a unified approach the notions,
notations and preliminary results on which our convergence analysis of the
proximal-projection method is based. It should be noted that, in Section 2,
some of the notions and results are presented in a more general setting than
strictly needed in the subsequent parts of the material. This is done so
because we hope to use the framework created in the current paper as a base
for a forthcoming study of methods of solving nonclassical variational
inequalities which are only tangentially approached here.

\section{Proximal Mappings, Relative Projections and Variational Inequalities%
}

In this section we present the notions, notations and results concerning
proximal projections and variational inequalities which are essential for
the convergence analysis of the proximal-projection method done in the
sequels.

\subsection{P\textbf{roximal mappings and relative projections}}

All over this paper we denote by $\mathcal{F}_{f}$ the set of proper, lower
semicontinuous, convex functions $\varphi :X\rightarrow (-\infty ,+\infty ]$
which satisfy the conditions that%
\begin{equation}
\mathrm{dom}\,\varphi \cap \mathrm{int\,dom}\,f\neq \varnothing ,
\label{Ff-i}
\end{equation}%
and%
\begin{equation}
\varphi _{f}:=\inf \left\{ \varphi (x):x\in \mathrm{dom}\,\varphi \cap 
\mathrm{dom}\,f\right\} >-\infty .  \label{Ff-ii}
\end{equation}%
With every $\varphi \in \mathcal{F}_{f}$ we associate the function \textrm{%
Env}$_{\varphi }^{f}:X^{\ast }\rightarrow (-\infty ,+\infty ]$ given by%
\begin{equation}
\mathrm{Env}_{\varphi }^{f}(\xi )=\inf \{\varphi (x)+W_{f}(\xi ,x):x\in X\}.
\label{Env}
\end{equation}%
This is a natural generalization of the notion of \textit{Moreau envelope
function} (see \cite[Definition 1.22]{RocWetts-Book}). By (\ref{Ff-i}) and (%
\ref{Ff-ii}) it results that the function $\mathrm{Env}_{\varphi }^{f}$ is
proper and $\mathrm{dom}\,$\textrm{Env}$_{\varphi }^{f}=\mathrm{dom}%
\,f^{\ast }.$ Another generalization of the \textit{Moreau envelope function 
}is the function $\mathrm{env}_{\varphi }^{f}:=\mathrm{Env}_{\varphi
}^{f}\circ \nabla f$ introduced and studied in \cite{BauComNoll-JointMin}.
Using Fenchel's duality theorem, it is easy to deduce that\textbf{\ }if%
\textit{\ }$\varphi \in \mathcal{F}_{f},$ then%
\begin{equation*}
\mathrm{Env}_{\varphi }^{f}(\xi )=f^{\ast }(\xi )-\left( \varphi +f\right)
^{\ast }(\xi )=f^{\ast }(\xi )-\left( \varphi ^{\ast }\square f^{\ast
}\right) (\xi ).
\end{equation*}%
where $\varphi ^{\ast }\square f^{\ast }$\ denotes the infimal convolution
of $\varphi ^{\ast }$\ and $f^{\ast }.$

The next result shows a way of generalizing the notion of Moreau proximal
mapping (in the sense given to this term in \cite{RocWetts-Book}) whose
study was initiated in \cite{Moreau1962}, \cite{Moreau1963}, \cite%
{Moreau1965} and further developed in \cite{RocProxPt}, \cite{RocAugLag}. As
we will make clear below, the generalization we propose here slightly
differs from the notion of $D_{f}$-proximal mapping introduced and studied
in \cite{BauBorCom-BregMon}. In fact, most of the next lemma can be also
deduced from \cite[Propositions 3.22 and 3.23]{BauBorCom-BregMon} due to the
equality (\ref{prox}) established below. We prefer to present it here with a
direct proof for sake of completeness.$\smallskip $

\textbf{Lemma 2.1: }\textit{Suppose that}\textbf{\ }$\varphi \in \mathcal{F}%
_{f}.$ \textit{For any }$\xi \in \mathrm{int\,dom}\,f^{\ast }$\textit{\
there exists a unique global minimizer, denoted }\textrm{Prox}$_{\varphi
}^{f}(\xi ),$ \textit{of the} \textit{function }$\varphi (\cdot )+W_{f}(\xi
,\cdot ).$ \textit{The vector\ }\textrm{Prox}$_{\varphi }^{f}(\xi )$ \textit{%
is contained in} $\mathrm{dom}\,\partial \varphi \cap \mathrm{int\,dom}\,f$ 
\textit{and we have}%
\begin{equation}
\mathrm{Prox}_{\varphi }^{f}(\xi )=\left[ \partial \left( \varphi +f\right) %
\right] ^{-1}(\xi )=\left( \partial \varphi +\nabla f\right) ^{-1}(\xi ).
\label{may}
\end{equation}

\textbf{Proof. }Let $\xi \in \mathrm{int\,dom}\,f^{\ast }.$ Then $\mathrm{Env%
}_{\varphi }^{f}(\xi )$ is finite and the function $f-\left\langle \xi
,\cdot \right\rangle $ is coercive (see \cite[Theorem 7A]{RocLevelSets} or 
\cite[Fact 3.1]{BauBorCom-Essential}), that is, its sublevel sets 
\begin{equation*}
\mathrm{lev}_{\leq }^{f}(\alpha ):=\left\{ x\in X:f(x)\leq \alpha \right\} ,
\end{equation*}%
are bounded for all $\alpha \in \mathbb{R}$. Consequently, the function $%
W_{f}(\xi ,\cdot )$ is coercive too. Let $\left\{ x^{k}\right\} _{k\in 
\mathbb{N}}$ be a sequence contained in $\mathrm{dom}\,\varphi \cap \mathrm{%
dom}\,f$ and such that%
\begin{equation*}
\lim_{k\rightarrow \infty }\left[ \varphi (x^{k})+W_{f}(\xi ,x^{k})\right] =%
\mathrm{Env}_{\varphi }^{f}(\xi ).
\end{equation*}%
The sequence $\left\{ \varphi (x^{k})+W_{f}(\xi ,x^{k})\right\} _{k\in 
\mathbb{N}}$ being convergent is also bounded. So, for some real number $M>0$
we have%
\begin{equation*}
W_{f}(\xi ,x^{k})\leq M-\varphi (x^{k})\leq M-\varphi _{f},
\end{equation*}%
showing that the sequence $\left\{ x^{k}\right\} _{k\in \mathbb{N}}$ is
contained in the sublevel set \textrm{lev}$_{\leq }^{\psi }(M-\varphi _{f})$
of the function $\psi :=W_{f}(\xi ,\cdot ).$ By the coercivity of $W_{f}(\xi
,\cdot )$ it follows that the sequence $\left\{ x^{k}\right\} _{k\in \mathbb{%
N}}$ is bounded. Since the space $X$ is reflexive, it results that $\left\{
x^{k}\right\} _{k\in \mathbb{N}}$ has a weakly convergent subsequence $%
\left\{ x^{i_{k}}\right\} _{k\in \mathbb{N}}.$ Let $\bar{x}=\,$w-$%
\lim_{k\rightarrow \infty }x^{i_{k}}.$ The functions $f$ and $\varphi $ are
sequentially weakly lower semicontinuous because they are lower
semicontinuous and convex. Hence, $\varphi (\cdot )+W_{f}(\xi ,\cdot )$ is
also sequentially weakly lower semicontinuous and, thus, we have%
\begin{equation*}
\varphi (\bar{x})+W_{f}(\xi ,\bar{x})\leq \underset{k\rightarrow \infty }{%
\lim \inf }\text{\thinspace }\left[ \varphi (x^{i_{k}})+W_{f}(\xi ,x^{i_{k}})%
\right] =\mathrm{Env}_{\varphi }^{f}(\xi )<+\infty .
\end{equation*}%
This implies that $\bar{x}\in \mathrm{dom}\,\varphi \cap \mathrm{dom}\,f$
and that $\bar{x}$ is a minimizer of $\varphi (\cdot )+W_{f}(\xi ,\cdot ).$

Suppose that $y$ is any minimizer of $\varphi (\cdot )+W_{f}(\xi ,\cdot ).$
Then $y$ is also a minimizer of $\varphi +f-$\thinspace $\xi .$ Therefore,
we have that $0\in \partial (\varphi +f-\xi )(y),$ that is, $\xi \in
\partial (\varphi +f)(y)$. The function $f$ is continuous on $\mathrm{int}\,%
\mathrm{dom}\,f$ (as being convex and lower semicontinuous). This and (\ref%
{Ff-i}) imply (see \cite{RocSums}) that $\partial (\varphi +f)(y)=\partial
\varphi (y)+\nabla f(y).$ Hence, 
\begin{equation}
\xi \in \partial \varphi (y)+\nabla f(y).  \label{key}
\end{equation}%
Since $\mathrm{dom}\,\nabla f=\mathrm{int\,dom}\,f$ (see (\ref{domf*'})),
this implies that 
\begin{equation*}
y\in \mathrm{dom}\,\partial \varphi \cap \mathrm{dom}\,\nabla f=\mathrm{dom}%
\,\partial \varphi \cap \mathrm{int\,dom}\,f.
\end{equation*}%
Hence, all minimizers of $\varphi (\cdot )+W_{f}(\xi ,\cdot )$ are contained
in $\mathrm{dom}\,\partial \varphi \cap \mathrm{int\,dom}\,f\subseteq 
\mathrm{dom}\,\varphi \cap \mathrm{int\,dom}\,f.$ The Legendre function $f$
is strictly convex on the convex subsets of $\mathrm{dom}\,\partial f$ and,
in particular, on the convex set $\mathrm{dom}\,\varphi \,\cap \,\mathrm{%
int\,dom}\,f=\mathrm{dom}\,\varphi \,\cap \,\mathrm{dom}\,\partial f$. Thus, 
$\varphi (\cdot )+W_{f}(\xi ,\cdot )$ is strictly convex on this set too.
Consequently, there is at most one minimizer of $\varphi (\cdot )+W_{f}(\xi
,\cdot )$ on the convex set $\mathrm{dom}\,\varphi \cap \mathrm{int\,dom}\,f$
and this proves that the minimizer $\bar{x}$ whose existence was established
above is unique. Formula (\ref{may}) follows from (\ref{key}) when $y=\bar{x}%
.\hfill \square \smallskip $

Lemma 2.1 ensures well definedness of the function 
\begin{equation}
{\mathrm{Prox}}_{\varphi }^{f}:\mathrm{int\,dom}\,f^{\ast }\rightarrow 
\mathrm{dom}\,\partial \varphi \cap \mathrm{int\,dom}\,f,\text{ }\xi
\rightarrow {\mathrm{Prox}}_{\varphi }^{f}(\xi )  \label{Prox}
\end{equation}%
when $\varphi \in \mathcal{F}_{f}.$ We call this function \textit{the
proximal mapping relative to} $f$ \textit{associated to} $\varphi .$ Well
definedness of the proximal mappings relative to the Legendre function $f$
can also be deduced from \cite[Theorem 3.18]{BauBorCom-BregMon} where well
definedness of the resolvent%
\begin{equation}
\mathrm{prox}_{\varphi }^{f}:={\mathrm{Prox}}_{\varphi }^{f}\circ \nabla f
\label{prox}
\end{equation}%
was established. In more particular circumstances for $f$ and $\varphi ,$
well definedness of ${\mathrm{Prox}}_{\varphi }^{f}$ was proved in \cite%
{Alber1993}, \cite{ButIusProxPt}, and in \cite{ButResSurvey}.

Let $E$ be a closed convex subset of $X$ satisfying 
\begin{equation}
E\cap \mathrm{int}\,\mathrm{dom}\,f\neq \varnothing .  \label{C}
\end{equation}%
Then the indicator function of the set $E,$ that is, the function $\iota
_{E}:X\rightarrow (-\infty ,+\infty ]$ defined by $\iota _{E}(x)=0$ if $x\in
E,$ and $\iota _{E}(x)=+\infty $, otherwise, is contained in $\mathcal{F}%
_{f}.$ The operator ${\mathrm{Prox}}_{i_{E}}^{f}$ is called \textit{%
projection onto }$E$\textit{\ relative to }$f$ (cf. \cite{ButResSurvey}) and
is denoted ${\mathrm{Proj}}_{E}^{f}$ in that follows. According to Lemma
2.1, we have%
\begin{equation*}
{\mathrm{Proj}}_{E}^{f}=(N_{E}+\nabla f)^{-1}
\end{equation*}%
where $N_{E}$ denotes the normal cone operator associated to the set $E.$
The operator 
\begin{equation}
\mathrm{proj}_{E}^{f}:=\mathrm{prox}_{\iota _{E}}^{f}={\mathrm{Proj}}%
_{_{E}}^{f}\circ \nabla f  \label{proj}
\end{equation}%
is exactly the \textit{Bregman projection onto }$E$\textit{\ relative to} $f$
whose importance in convex optimization was first emphasized in \cite%
{Bregman-Relaxation}. To see that it is sufficient to recall that the 
\textit{Bregman distance} $D_{f}:X\times \mathrm{int}\,\mathrm{dom}%
\,f\rightarrow (-\infty ,+\infty ]$ is the function defined by%
\begin{equation}
D_{f}(y,x):=f(y)-f(x)-\left\langle \nabla f(x),y-x\right\rangle
=W_{f}(\nabla f(x),y),  \label{Df}
\end{equation}%
with $\mathrm{dom}\,D_{f}=\left( \mathrm{dom}\,f\right) \times \left( 
\mathrm{int\,dom}\,f\right) .$\medskip

\subsection{Variational inequalities}

There is an intimate connection between proximal mappings and variational
inequalities. It is based on the following result which extends the
variational characterization of ${\mathrm{Proj}}_{E}^{f}$ given in \cite%
{ButResSurvey} to a variational characterization of ${\mathrm{Prox}}%
_{\varphi }^{f}.\smallskip $

\textbf{Lemma 2.2. }\textit{Suppose that}\textbf{\ }$\varphi \in \mathcal{F}%
_{f}$\textit{\ and }$\xi \in \mathrm{int\,dom}\,f^{\ast }.$\textit{\ If }$%
\hat{x}\in \,\mathrm{dom}\,\,\partial \varphi \cap \,\mathrm{int}\,\,\,%
\mathrm{dom}\,\,f$\textit{\ then the following conditions are equivalent:}

($a$) $\hat{x}=${\textrm{Prox}}$_{\varphi }^{f}(\xi );$

($b$) $\hat{x}$ \textit{is a solution of the variational inequality} 
\begin{equation}
\left\langle \xi -\nabla f(x),y-x\right\rangle \leq \varphi (y)-\varphi (x),%
\text{ }\forall y\in \mathrm{dom}\,\varphi \cap \mathrm{dom}\,f.
\label{proxvi}
\end{equation}

($c$) $\hat{x}$ \textit{is a solution of the variational inequality}%
\begin{equation*}
W_{f}(\xi ,x)+W_{f}(\nabla f(x),y)-W_{f}(\xi ,y)\leq \varphi (y)-\varphi (x),%
\text{ }\forall y\in \mathrm{dom}\,\varphi \cap \mathrm{dom}\,f.
\end{equation*}

\textbf{Proof: }\frame{($a$)$\Rightarrow $($b$)} Suppose that $\hat{x}=${%
\textrm{Prox}}$_{\varphi }^{f}(\xi ).$ Take $y\in \mathrm{dom}\,\varphi \cap 
\mathrm{dom}\,f$ and $t\in (0,1).$ Then $(1-t)\hat{x}+ty\in \mathrm{dom}%
\,\varphi \cap \mathrm{dom}\,f$ and we have%
\begin{equation*}
\varphi (\hat{x})+W_{f}(\xi ,\hat{x})\leq \varphi ((1-t)\hat{x}%
+ty)+W_{f}(\xi ,(1-t)\hat{x}+ty),
\end{equation*}%
that is%
\begin{equation}
\varphi ((1-t)\hat{x}+ty)-\varphi (\hat{x})\geq W_{f}(\xi ,\hat{x}%
)-W_{f}(\xi ,(1-t)\hat{x}+ty).  \label{pips}
\end{equation}%
Since $\hat{x}\in \mathrm{int\,dom}\,f,$ there exists $t_{0}\in (0,1)$ such
that for any $t\in (0,t_{0})$ we have that $(1-t)\hat{x}+ty\in \mathrm{%
int\,dom}\,f.$ Consequently, for any $t\in (0,t_{0})$ the function $%
W_{f}(\xi ,\cdot )$ is differentiable at $(1-t)\hat{x}+ty.$ Clearly, we also
have%
\begin{equation*}
\nabla W_{f}(\xi ,\cdot )(u)=\nabla f(u)-\xi ,\text{ }\forall u\in \mathrm{%
int\,dom}\,f.
\end{equation*}%
Therefore, by the convexity of $W_{f}(\xi ,\cdot )$ and (\ref{pips}), we
deduce%
\begin{eqnarray*}
t^{-1}\left[ \varphi ((1-t)\hat{x}+ty)-\varphi (\hat{x})\right] &\geq
&\left\langle \nabla W_{f}(\xi ,\cdot )((1-t)\hat{x}+ty),\hat{x}%
-y\right\rangle \\
&=&\left\langle \nabla f((1-t)\hat{x}+ty)-\xi ,\hat{x}-y\right\rangle
\end{eqnarray*}%
for any $t\in (0,t_{0}).$ The function $\nabla f(\cdot )$ is norm to weak
continuous (see, for instance, \cite[Proposition 2.8]{PhelpsBook}). Hence,
letting $t\rightarrow 0^{+}$ in the last inequality we get%
\begin{equation*}
\varphi ^{\circ }(\hat{x};y-\hat{x})\geq \left\langle \nabla f(\hat{x})-\xi ,%
\hat{x}-y\right\rangle ,
\end{equation*}%
where $\varphi ^{\circ }$ stands for the right-hand side derivative of $%
\varphi $, that is,%
\begin{equation*}
\varphi ^{\circ }(x;d)=\lim_{s\rightarrow 0^{+}}\frac{\varphi (x+sd)-\varphi
(x)}{s}.
\end{equation*}%
Taking into account that%
\begin{equation*}
\varphi (y)-\varphi (\hat{x})\geq \varphi ^{\circ }(\hat{x};y-\hat{x})
\end{equation*}%
we obtain (\ref{proxvi}).

\frame{($b$)$\Rightarrow $($a$)} Suppose that%
\begin{equation*}
\left\langle \xi -\nabla f(\hat{x}),y-\hat{x}\right\rangle \leq \varphi
(y)-\varphi (\hat{x}),\text{ }\forall y\in \mathrm{dom}\,\varphi \cap 
\mathrm{dom}\,f.
\end{equation*}%
Observe that 
\begin{equation}
\nabla W_{f}(\xi ,\cdot )=\nabla f(\cdot )-\xi .  \label{diffWf}
\end{equation}%
Then, by the convexity of $W_{f}(\xi ,\cdot )$ and (\ref{diffWf}), for any $%
y\in \mathrm{dom}\,\varphi \cap \mathrm{dom}\,f$ we have that%
\begin{eqnarray*}
W_{f}(\,\xi ,y)-W_{f}(\,\xi ,\hat{x}) &\geq &\left\langle \nabla W_{f}(\,\xi
,\cdot )(\hat{x}),y-\hat{x}\right\rangle \\
&=&\left\langle \nabla f(\hat{x})-\,\xi ,y-\hat{x}\right\rangle \\
&\geq &\varphi (\hat{x})-\varphi (y).
\end{eqnarray*}%
This shows that $\hat{x}=${\textrm{Prox}}$_{\varphi }^{f}($\thinspace $\xi
). $ The equivalence $\frame{($b$)$\Leftrightarrow $($c$)}$ results
immediately by observing that%
\begin{equation*}
W_{f}(\,\xi ,x)+W_{f}(\nabla f(x),y)-W_{f}(\,\xi ,y)=\left\langle \xi
-\nabla f(x),y-x\right\rangle ,
\end{equation*}%
whenever $y\in \mathrm{dom}\,f,$ \thinspace $\xi \in \mathrm{int}\,\mathrm{%
dom}\,f^{\ast }$ and $x\in \mathrm{int}\,\mathrm{dom}\,f.\hfill \square
\medskip $

A consequence of Lemma 2.2 is the following generalization of the
variational characterization of the Bregman projections originally given in 
\cite{AlbBut}.\smallskip

\textbf{Corollary 2.1. }\textit{Let }$x\in \mathrm{int}\,\,\mathrm{dom}\,f$ 
\textit{and let} $E$ \textit{be a nonempty, closed and convex set such that }%
$E\cap \,\mathrm{int}\,\,\mathrm{dom}\,f\neq \varnothing $. \textit{If} $%
\hat{x}\in E,$ \textit{then the following conditions are equivalent:}

($i$)\textit{\ The vector} $\hat{x}$ \textit{is the Bregman projection of} $%
x $ \textit{onto} $E$ \textit{with respect to} $f$;

($ii$)\textit{\ The vector }$\hat{x}$\textit{\ is the unique solution of the
variational inequality} 
\begin{equation*}
\left\langle \nabla f(x)-\nabla f(z),z-y\right\rangle \geq 0,\quad \forall
y\in E;
\end{equation*}

($iii$) \textit{The vector }$\hat{x}$\textit{\ is the unique solution of the
variational inequality} 
\begin{equation*}
D_{f}(y,z)+D_{f}(z,x)\leq D_{f}(y,x),\quad \forall y\in K.
\end{equation*}

Now we are in position to establish the connection between the proximal
mappings and a class of variational inequalities. It extends similar results
known to hold in less general settings (see, for instance,\cite{Alber1993}
and \cite[Proposition 1.5.8]{FacPan}). The variational inequality we
consider here is%
\begin{equation}
\text{Find }x\in \mathrm{int}\,\mathrm{dom}\,f\text{ such that}
\label{genVI}
\end{equation}%
\begin{equation*}
\exists \,\xi \in Bx:\left[ \left\langle \,\xi ,y-x\right\rangle \geq
\varphi (x)-\varphi (y),\text{ }\forall y\in \mathrm{dom}\,f\right] ,
\end{equation*}%
where $\varphi \in \mathcal{F}_{f}$ and $B:X\rightarrow 2^{X^{\ast }}$%
\textit{\ }is an operator which satisfies the condition%
\begin{equation}
\varnothing \neq \mathrm{dom}\,B\cap \mathrm{dom}\,\varphi \cap \mathrm{int}%
\,\mathrm{dom}\,f\text{ and }\mathrm{ran}\,(\nabla f-B)\subseteq \mathrm{int}%
\,\mathrm{dom}\,f^{\ast }.  \label{compat}
\end{equation}%
Condition (\ref{compat}) guarantees that the operator 
\begin{equation*}
\mathrm{Prox}_{\varphi }^{f}\left( \nabla f-B\right) :=\mathrm{Prox}%
_{\varphi }^{f}\circ \left( \nabla f-B\right)
\end{equation*}%
is well defined. Therefore, the following statement makes sense.\smallskip

\textbf{Lemma 2.3. }\textit{Let }$\varphi \in \mathcal{F}_{f}$ \textit{and} $%
\hat{x}\in \,\mathrm{dom}\,\partial \varphi \cap \mathrm{int}\,\mathrm{dom}%
\,f$\textit{. Suppose that }$B:X\rightarrow 2^{X^{\ast }}$\textit{\ is an
operator which satisfies }(\ref{compat}). \textit{Then } $\hat{x}$ \textit{%
is a solution of} \textit{the variational inequality }(\ref{genVI}) \textit{%
if and only if} \textit{it} \textit{is a fixed point of the operator }%
\textrm{Prox}$_{\varphi }^{f}\left( \nabla f-B\right) .\smallskip $

\textbf{Proof:} Note that $\hat{x}$ is a solution of (\ref{genVI}) if and
only if there exists $\xi \in B\hat{x}$ such that%
\begin{equation}
\left\langle (\nabla f(\hat{x})-\,\xi )-\nabla f(\hat{x}),y-\hat{x}%
\right\rangle \leq \varphi (y)-\varphi (\hat{x}),\text{ }\forall y\in 
\mathrm{dom}\,\varphi \cap \mathrm{dom}\,f.  \label{jump}
\end{equation}%
According to Lemma 2.2, this is equivalent to $\hat{x}=\mathrm{Prox}%
_{\varphi }^{f}\left( \nabla f(\hat{x})-\,\xi \right) $ for some $\xi \in B%
\hat{x}$ which, in turn, is equivalent to $\hat{x}\in \mathrm{Prox}_{\varphi
}^{f}\left( \nabla f(\hat{x})-B\hat{x}\right) ,$ i.e., to the condition that 
$\hat{x}$ is a fixed point of $\mathrm{Prox}_{\varphi }^{f}\left( \nabla
f-B\right) .\hfill \square $\medskip

Let $B:X\rightarrow 2^{X^{\ast }}$ be an operator and suppose that the
closed convex subset $C$ of $X$ satisfies%
\begin{equation}
\varnothing \neq \mathrm{dom}\,B\cap C\cap \mathrm{int}\,\mathrm{dom}\,f%
\text{ and }\mathrm{ran}\,(\nabla f-B)\subseteq \mathrm{int}\,\mathrm{dom}%
\,f^{\ast }.  \label{compatvi}
\end{equation}%
Note that if $\varphi :=\iota _{C},$ then the variational inequality (\ref%
{genVI}) is exactly a \textit{classical variational inequality}%
\begin{equation}
\text{Find }x\in C\cap \mathrm{int}\,\mathrm{dom}\,f\text{ such that}
\label{vi}
\end{equation}%
\begin{equation*}
\exists \,\xi \in Bx:\left[ \left\langle \,\xi ,y-x\right\rangle \geq 0,%
\text{ }\forall y\in C\cap \mathrm{dom}\,f\right] .
\end{equation*}%
Applying Lemma 2.3 and (\ref{f'f*}) in this case, we re-discover the
following known result (cf. \cite{alberGenproj}):\smallskip

\textbf{Lemma 2.4. }\textit{Suppose that the condition }(\ref{compatvi})%
\textit{\ is satisfied and that} $\hat{x}\in C\cap \mathrm{int}\,\mathrm{dom}%
\,f.$ \textit{Then the following statements are equivalent:}

($a$) \textit{The vector }$\hat{x}$\textit{\ is a solution of the classical
variational inequality} (\ref{vi})\textit{;}

($b$) \textit{The vector }$\hat{x}$ \textit{is a fixed point of the operator 
}\textrm{Proj}$_{C}^{f}\left( \nabla f-B\right) $\textit{;}

($c$) \textit{The vector }$\hat{x}$\textit{\ is a zero for the operator }$%
V[B;C;f]:X\rightarrow 2^{X^{\ast }}$\textit{\ given by}%
\begin{equation}
V[B;C;f]:=\nabla f-\nabla f\circ \mathrm{Proj}_{C}^{f}\left( \nabla
f-B\right) .  \label{fakeB}
\end{equation}%
\smallskip

Lemma 2.3 and Lemma 2.4 provided the initial motivation for this research.
Observe that Lemma 2.4 reduces the problem of finding a solution for the
classical variational inequality (\ref{vi}) to the problem of finding a
fixed point for the operator \textrm{Proj}$_{C}^{f}\left( \nabla f-B\right) $%
. It is well known that, in many instances, by iterating an operator
starting from initial points located in its definition domain, one produces
sequences which converge to fixed points of the operator. This suggests that
for solving the classical variational inequality (\ref{vi}) we would have to
produce sequences defined by the iterative rule 
\begin{equation}
x^{k+1}\in \mathrm{Proj}_{C}^{f}(\nabla f(x^{k})-Bx^{k})  \label{star1}
\end{equation}%
in hope that such sequences will converge to fixed points of \textrm{Proj}$%
_{C}^{f}\left( \nabla f-B\right) .$ Note that any fixed point of the
operator \textrm{Proj}$_{C}^{f}\left( \nabla f-B\right) $ is a zero for the
operator $V[B;C;f]$ given by (\ref{fakeB}) and conversely. According to (\ref%
{proj}), we have that%
\begin{eqnarray}
\mathrm{Proj}_{C}^{f}\left( \nabla f-V[B;C;f]\right) &=&\mathrm{Proj}%
_{C}^{f}\circ \nabla f\circ \mathrm{Proj}_{C}^{f}\circ \left( \nabla
f-B\right)  \label{star2} \\
&=&\mathrm{proj}_{C}^{f}\circ \mathrm{proj}_{C}^{f}\circ \left( \nabla
f\right) ^{-1}\circ \left( \nabla f-B\right)  \notag \\
&=&\mathrm{proj}_{C}^{f}\circ \nabla f^{\ast }\circ \left( \nabla f-B\right)
\notag \\
&=&\mathrm{Proj}_{C}^{f}\left( \nabla f-B\right) .  \notag
\end{eqnarray}%
Thus, we are naturally led to the question of whether, and in which
conditions, the sequences generated according to the rule $x^{k+1}\in 
\mathrm{Proj}_{C}^{f}(\nabla f(x^{k})-V[B;C;f]x^{k}),$ which are the same
(see (\ref{star2})) as the sequences generated according to rule (\ref{star1}%
), converge to zeros of the operator $V[B;C;f].$ This is, in fact, a
particular instance of the more general question of whether, and in which
conditions, the proximal-projection method (\ref{alg}) approximates zeros of
a given operator $V[B;C;f],$ provided that such zeros exist. In the sequels
we present answers to this question. It is interesting to observe that by
focusing in our convergence analysis on conditions concerning the operator $%
V[B;C;f]$ instead of $B$ we do not mean that computing values of \textrm{Proj%
}$_{C}^{f}\left( \nabla f-V[B;C;f]\right) $ is easier than computing values
of the same operator via the formula \textrm{Proj}$_{C}^{f}\left( \nabla
f-B\right) .$ However, from a theoretical point of view, the operator $%
V[B;C;f]$ associated to $B$ via formula (\ref{fakeB}) may happen to be
better conditioned than $B$ for a convergence analysis of the
proximal-projection method. This aspect can be clearly seen after a careful
dissection of the considerations which lead to our main convergence results
presented in this paper. It should be taken into account that the operator $%
B $ may have not zeros in $C$ even if the operator $V[B;C;f],$ associated to 
$B $ by (\ref{fakeB}), has. For example, take $X=\mathbb{R}$, $f(x)=%
\textstyle%
\frac{1}{2}%
x^{2},$ $C=[1,2]$ and $Bx=x$ for all $x\in X.$ Then $V[B;C;f]x=x-1$ vanishes
at $x=1,$ but $B$ does not have any zero in $C.$

\section{$D_{f}$-nonexpansivity poles, $D_{f}$-coercivity and $D_{f}$%
-firmness of operators in Banach spaces}

In this section we introduce the notion of $D_{f}$-coercivity for operators
from $X$ to $2^{X^{\ast }}.$ We clarify how this notion is related with the
notions of $D_{f}$-nonexpansivity pole introduced in \cite{ButIus-Book} and
of $D_{f}$-firm operator introduced in \cite{BauBorCom-BregMon}. Using these
relations we show that the proximal point method with Bregman distances can
be seen as a particular instance of the proximal-projection method applied
to a $D_{f}$-coercive operator.

\subsection{$D_{f}$-nonexpansivity poles}

In that follows, to the function $f$ described in Section 1 and to any
operator $A:X\rightarrow 2^{X^{\ast }}$ we associate the operator $%
A^{f}:X\rightarrow 2^{X}$ given by%
\begin{equation}
A^{f}:=\nabla f^{\ast }\circ \left( \nabla f-A\right) .  \label{Afa}
\end{equation}%
We call this operator the $D_{f}$\textit{-antiresolvent of }$A$. Observe
that 
\begin{equation}
\mathrm{dom}\,A^{f}\subseteq \mathrm{dom}\,A\cap \mathrm{int}\,\mathrm{dom}%
\,f\text{ \quad and\quad\ }\mathrm{ran}\,A^{f}\subseteq \mathrm{int}\,%
\mathrm{dom}\,f,  \label{Afa1}
\end{equation}%
and that, if $x\in \mathrm{int}\,\mathrm{dom}\,f,$ then $0^{\ast }\in Ax$ if
and only if $x\in \mathrm{Fix\,}A^{f},$ where $\mathrm{Fix\,}A^{f}$ denotes
the set of fixed points of $A^{f}.$ Therefore, the (possibly empty) set of
solutions of the Problem 1.1 situated in $\mathrm{int}\,\mathrm{dom}\,f$,
denoted $\mathcal{S}_{f}(A,C),$ is exactly 
\begin{equation}
\mathcal{S}_{f}(A,C)=C\cap \mathrm{Fix\,}A^{f}.  \label{solopec}
\end{equation}%
We are going to prove that, for operators $A$ which are $D_{f}$-coercive,
the set $\mathcal{S}_{f}(A,C)$ is exactly the set of $D_{f}$-nonexpansivity
poles of $A^{f}$ over the set $C$ and this fact will be later used in our
convergence analysis of the proximal-projection method. To this end, recall
the following notion.\smallskip

\textbf{Definition 3.1 }(cf. \cite{ButIus-Book}) Let $T:X\rightarrow 2^{X}$
be an operator and let $Y$ be a subset of $X$ such that%
\begin{equation}
\varnothing \neq T(Y\cap \mathrm{int}\,\mathrm{dom}\,f)\subseteq \mathrm{int}%
\,\mathrm{dom}\,f.  \label{tonton1}
\end{equation}%
The vector $z\in X$ is called a $D_{f}$-\textit{nonexpansivity pole of }$T$%
\textit{\ over }$Y$ if the following conditions are satisfied:%
\begin{equation}
z\in Y\cap \mathrm{int}\,\mathrm{dom}\,f,  \label{trombon}
\end{equation}%
\begin{equation}
(x\in Y\cap \mathrm{int}\,\mathrm{dom}\,f\text{\quad and\quad }u\in
Tx)\Rightarrow \left\langle \nabla f(x)-\nabla f(u),z-u\right\rangle \leq 0.
\label{tonton2}
\end{equation}%
We denote by \textrm{Nexp}$_{Y}^{f}\,T$ the set of $D_{f}$-nonexpansivity
poles of $T$ over $Y.$\smallskip

Operators having $D_{f}$-nonexpansivity poles were termed totally
nonexpansive operators in \cite{ButIus-Book}. Operators $T$ such that $%
\mathrm{ran}\,T\subseteq \mathrm{dom}\,T=\mathrm{int}\,\mathrm{dom}\,f$ and
having \textrm{Nexp}$_{\mathrm{int}\,\mathrm{dom}\,f}^{f}\,T\supseteq 
\mathrm{Fix}\,T$ were called $\mathcal{B}$\textit{-class operators} in \cite%
{BauBorCom-Essential} and \cite{BauBorCom-BregMon}. $\mathcal{B}$-class
operators necessarily have \textrm{Nexp}$_{\mathrm{int}\,\mathrm{dom}%
\,f}^{f}\,T=\mathrm{Fix}\,T$ (cf. \cite[Proposition 3.3]{BauBorCom-BregMon}%
). However, not every operator having $D_{f}$-nonexpansivity poles over some
subset $Y$ of $X$ is $\mathcal{B}$-class. For example, the operator $%
Tx=\{x^{2}\}$ when $X=\mathbb{R}$, $f=\textstyle%
\frac{1}{2}%
\left\vert \cdot \right\vert ^{2}$ and $Y=\mathrm{int}\,\mathrm{dom}\,f=%
\mathbb{R}$ has $z\in \,$\textrm{Nexp}$_{\mathrm{int}\,\mathrm{dom}%
\,f}^{f}\,T$ if and only if%
\begin{equation*}
x(1-x)(z-x^{2})\leq 0,\quad \forall x\in \mathbb{R}\text{,}
\end{equation*}%
and this last inequality can not hold because, irrespective of $z,$ we have $%
\lim_{x\rightarrow \infty }x(1-x)(z-x^{2})=\infty .$ Hence, \textrm{Nexp}$_{%
\mathrm{int}\,\mathrm{dom}\,f}^{f}\,T=\varnothing .$ In spite of that, $%
\mathrm{Fix}\,T=\{0,1\}$ and, if $Y=[0,1],$ then (\ref{tonton2}) holds for $%
z=0$ only, i.e., \textrm{Nexp}$_{[0,1]}^{f}\,T=\{0\}\neq \mathrm{Fix}\,T.$

The following lemma summarizes several properties of operators having
nonexpansivity poles which are used in this work.\smallskip

\textbf{Lemma 3.1. }\textit{Let the operator }$T:X\rightarrow 2^{X}$\textit{%
\ and the set }$Y\subseteq X$\textit{\ be such that condition} (\ref{tonton1}%
) \textit{holds. Then the following statements are true:}

($a$) \textit{The (possibly empty) set }\textrm{Nexp}$_{Y}^{f}\,T$\textit{\
is convex and closed when }$Y\subseteq \mathrm{int}\,\mathrm{dom}\,f$\textit{%
\ is convex and closed;}

($b$) \textit{A vector} $z\in Y\cap \mathrm{int}\,\mathrm{dom}\,f$ \textit{%
is a }$D_{f}$\textit{-nonexpansivity pole of }$T$\textit{\ over }$Y$\textit{%
\ if and only if} 
\begin{equation}
(x\in Y\cap \mathrm{int}\,\mathrm{dom}\,f\text{\quad and\quad }u\in
Tx)\Rightarrow D_{f}(z,u)+D_{f}(u,x)\leq D_{f}(z,x).  \label{tn}
\end{equation}

($c$) $\mathrm{Nexp}_{Y}^{f}\,T\subseteq \,$\textrm{Fix}$\,T$ \textit{and }$%
T $\textit{\ is single-valued at any }$D_{f}$\textit{-nonexpansivity
pole.\medskip }

\textbf{Proof. }Statement ($a$) results from the fact that the function $%
z\rightarrow $\linebreak $\left\langle \nabla f(x)-\nabla
f(u),z-u\right\rangle $ is linear and continuous of $z.$ Statement ($b$)
follows from (\ref{tonton2}) and (\ref{Df}). Now, by taking in (\ref{tn}) $%
x=z\in \mathrm{Nexp}_{Y}^{f}\,T$ one gets 
\begin{equation}
D_{f}(z,u)=D_{f}(u,z)=0,\text{ }\forall u\in Tz.  \label{pil}
\end{equation}%
By (\ref{tonton1}), if $u\in Tz,$ then $u\in \mathrm{int}\,\mathrm{dom}\,f.$
The function $f$ being essentially strictly convex is strictly convex on $%
\mathrm{int}\,\mathrm{dom}\,f.$ Hence, the equalities in (\ref{pil}) can not
hold unless $u=z$ (cf. \cite[Proposition 1.1.4]{ButIus-Book}). In other
words, we have the following implication%
\begin{equation}
z\in \mathrm{Nexp}_{Y}^{f}\,T\Rightarrow Tz=\{z\}.  \label{fp-np}
\end{equation}%
It shows that $\mathrm{Nexp}_{Y}^{f}\,T\subseteq \,$\textrm{Fix}$\,T$ and
that $T$ is single-valued at $D_{f}$-nonexpansivity poles.$\hfill \square $

\subsection{$D_{f}$-coercivity and $D_{f}$-firmness}

The notion of $D_{f}$-coercive operator, introduced in this section, and the
notion of $D_{f}$-firm operator, originally introduced in \cite[Definition
3.4]{BauBorCom-BregMon}, are generalizations of the notion of firmly
nonexpansive operator in a Hilbert space. Recall (cf. \cite[pp. 41-42]%
{GoeRei}) that if $X$ is a Hilbert space (which we always identify with its
dual $X^{\ast }$), then an operator $A:X\rightarrow X$ is firmly
nonexpansive on a subset $Y$ of $X$ if and only if 
\begin{equation}
\left\langle Ax-Ay,x-y\right\rangle \geq \left\Vert Ax-Ay\right\Vert ^{2},%
\text{ }\forall x,y\in Y.  \label{f-n}
\end{equation}%
It can be easily seen from the definitions given below that if $X$ is a
Hilbert space and if $f=\textstyle%
\frac{1}{2}%
\left\Vert \cdot \right\Vert ^{2}$, then the operator $A$ is $D_{f}$%
-coercive on its domain if and only if it is firmly nonexpansive on its
domain and this happens if and only if $A$ is $D_{f}$-firm.

Returning to the general context in which $X$ and $f$ are as described in
Section 1, we introduce the following notion.\smallskip

\textbf{Definition 3.2.} Let $Y$ be a subset of the space $X.$ The operator $%
A:X\rightarrow 2^{X^{\ast }}$ is called $D_{f}$-\textit{coercive} \textit{on}
\textit{the set} $Y$ if 
\begin{equation}
Y\cap \left( \mathrm{dom}\,A\right) \cap \left( \mathrm{int}\,\mathrm{dom}%
\,f\right) \neq \varnothing  \label{YdomAdomf}
\end{equation}%
and 
\begin{equation}
\left. 
\begin{array}{c}
x,y\in Y\cap \mathrm{int}\,\mathrm{dom}\,f \\ 
\,\xi \in Ax\text{ and }\eta \in Ay%
\end{array}%
\right\} \Rightarrow \left\langle \,\xi -\eta ,\nabla f^{\ast }\left( \nabla
f(x)-\,\xi \right) -\nabla f^{\ast }\left( \nabla f(y)-\eta \right)
\right\rangle \geq 0.  \label{InvMon}
\end{equation}

Operators satisfying a somewhat less restrictive condition than (\ref{InvMon}%
) were studied in \cite[Section 5]{ButResSurvey} under the name of
inverse-monotone operators relative to $f.$ In general, an operator $A$
(even in a Hilbert space provided that $f$ is not the function $f=\textstyle%
\frac{1}{2}%
\left\Vert \cdot \right\Vert ^{2}$) does not have to satisfy (\ref{f-n}) in
order to be $D_{f}$-coercive on $Y$. For instance, if the function $f$ has a
minimizer in $\mathrm{int}\,\mathrm{dom}\,f$ (i.e., if the equation $\nabla
f(x)=0$ has a solution), and if $\alpha \in (0,1),$ then the operator $%
A=\alpha \nabla f$ is $D_{f}$-coercive on $Y=\mathrm{dom}\,A=\mathrm{int}\,%
\mathrm{dom}\,f$ without necessarily satisfying (\ref{f-n}) on $Y=\mathrm{int%
}\,\mathrm{dom}\,f$. Indeed, in this case, if $x\in \mathrm{int}\,\mathrm{dom%
}\,f$ then $\left\{ \nabla f(x),0^{\ast }\right\} \subset \mathrm{ran}%
\,\nabla f=\mathrm{int}\,\mathrm{dom}\,f^{\ast }$ and, due to the convexity
of $\mathrm{int}\,\mathrm{dom}\,f^{\ast },$ we have that%
\begin{equation*}
(1-\alpha )\nabla f(x)=(1-\alpha )\nabla f(x)+\alpha 0^{\ast }\in \mathrm{int%
}\,\mathrm{dom}\,f^{\ast }=\mathrm{dom}\,\nabla f^{\ast },
\end{equation*}%
and, consequently, the operator%
\begin{equation*}
A^{f}x=\nabla f^{\ast }((1-\alpha )\nabla f(x))
\end{equation*}%
has $\mathrm{dom}\,A^{f}=\mathrm{int}\,\mathrm{dom}\,f$. If $x,y\in \mathrm{%
int}\,\mathrm{dom}\,f,$ and if $\beta =1-\alpha ,$ then, by the monotonicity
of $\nabla f^{\ast },$ we deduce that%
\begin{equation*}
\left\langle \,Ax-Ay,\nabla f^{\ast }\left( \nabla f(x)-Ax\right) -\nabla
f^{\ast }\left( \nabla f(y)-Ay\right) \right\rangle =
\end{equation*}%
\begin{equation*}
\alpha \beta ^{-1}\left\langle \beta \nabla f(x)-\beta \nabla f(y),\nabla
f^{\ast }(\beta \nabla f(x))-\nabla f^{\ast }(\beta \nabla
f(y))\right\rangle \geq 0,
\end{equation*}%
i.e., the operator $A=\alpha \nabla f$ is $D_{f}$-coercive on $\mathrm{int}\,%
\mathrm{dom}\,f.$ However, the operator $A=\alpha \nabla f$ does not have to
be firmly nonexpansive on its domain even if $X$ is a Hilbert space. For
example, take in the considerations above $X=L^{2}[0,1],$ $f=\textstyle%
\frac{1}{3}%
\left\Vert \cdot \right\Vert ^{3}$ and $\alpha \in (1/4,1).$ Then $\nabla
f(x)=\left\Vert x\right\Vert x$ and an easy verification shows that (\ref%
{f-n}) does not hold for any $x,y\in X.$ For instance, (\ref{f-n}) is
violated when $x=2\alpha ^{-1/2}$ and $y=0$.

We are going to show that there are strong connections between the $D_{f}$%
-coercivity of the operator $A$ and the $D_{f}$-firmness of its $D_{f}$%
-antiresolvent $A^{f}$. For this purpose we recall the following:\smallskip

\textbf{Definition 3.3.} (cf. \cite[Definition 3.4]{BauBorCom-BregMon}) An
operator $T:X\rightarrow 2^{X}$ is called $D_{f}$\textit{-firm }if it
satisfies the conditions 
\begin{equation}
\varnothing \neq \mathrm{dom}\,T\cup \mathrm{ran}\,T\subseteq \mathrm{int}\,%
\mathrm{dom}\,f  \label{DomT}
\end{equation}%
and%
\begin{equation}
u\in Tx\text{ and }v\in Ty\Rightarrow \left\langle \nabla f(u)-\nabla
f(v),u-v\right\rangle \leq \left\langle \nabla f(x)-\nabla
f(y),u-v\right\rangle .  \label{Dfirm}
\end{equation}

We start with the following result which summarizes some basic properties of 
$D_{f}$-coercive operators.\smallskip

\textbf{Lemma 3.2. }\textit{Let }$A:X\rightarrow 2^{X^{\ast }}$ \textit{be
an operator and let }$Y$\textit{\ be a subset of }$X$\textit{\ which
satisfies }(\ref{YdomAdomf}). \textit{The following statements are true:}

($a$) \textit{The operator }$A$\textit{\ is }$D_{f}$\textit{-coercive on }$Y$%
\textit{\ if and only it satisfies the following condition for any} $x,y\in
Y\cap \mathrm{int}\,\mathrm{dom}\,f:$%
\begin{equation*}
\left. 
\begin{array}{c}
u\in A^{f}x \\ 
v\in A^{f}y%
\end{array}%
\right\} \Rightarrow D_{f}(u,v)+D_{f}(v,u)+D_{f}(u,x)+D_{f}(v,y)\leq
D_{f}(v,x)+D_{f}(u,y);
\end{equation*}

($b$) \textit{If }$A$\textit{\ is }$D_{f}$\textit{-coercive on }$Y,$\textit{%
\ then }$A$\textit{\ is }$D_{f}$\textit{-coercive on any subset of }$Y$%
\textit{\ which intersects} $\mathrm{dom}\,A\cap \mathrm{int}\,\mathrm{dom}%
\,f$\textit{;}

($c$) \textit{The operator }$A$\textit{\ is }$D_{f}$\textit{-coercive on its
domain if and only if its }$D_{f}$\textit{-antiresolvent }$A^{f}$\textit{\
is }$D_{f}$\textit{-firm;}

($d$) \textit{If }$A$\textit{\ is }$D_{f}$\textit{-coercive on its domain,
then }$A^{f}$\textit{\ is single valued and all its fixed points are }$D_{f}$%
-\textit{nonexpansivity poles on }$\mathrm{int}\,\mathrm{dom}\,f$\textit{%
.\medskip }

\textbf{Proof. }Statements ($a$), ($b$) and ($c$) result from (\ref{Afa}), (%
\ref{Afa1}), (\ref{DomT}), (\ref{Dfirm}) and (\ref{Df}). Single valuedness
of $A^{f}$ in statement\textbf{\ }($d$) is a consequence of ($c$) and of 
\cite[Proposition 3.5(iii)]{BauBorCom-BregMon}. Letting $y=z\in \mathrm{Fix}%
\,A^{f}$ in the inequality of ($a$)$,$ and taking into account the single
valuedness of $A^{f},$ one obtains that $z$ satisfies (\ref{tn}) for $%
T=A^{f} $.$\hfill \square \medskip $

Whenever the operator $A$ involved in Problem 1.1 is $D_{f}$-coercive on the
domain $C$ of the problem, we have%
\begin{equation}
\mathcal{S}_{f}(A,C)=C\cap \mathrm{Fix}\,A^{f}=\mathrm{Nexp}_{C}^{f}\,A^{f}.
\label{solcoer}
\end{equation}%
This immediately follows from the next result.\smallskip

\textbf{Lemma 3.3. }\textit{The following statements are true:}

($a$) \textit{If} $z\in \mathrm{Nexp}_{C}^{f}\,A^{f},$ \textit{then} $%
Az=\{0^{\ast }\};$

($b$) \textit{If\ the operator }$A$\textit{\ is }$D_{f}$-\textit{coercive on 
}$C$\textit{\ and }$z\in \mathrm{int}\,\mathrm{dom}\,f$\textit{\ is a
solution of Problem 1.1, then} $z\in \mathrm{Nexp}_{C}^{f}\,A^{f}.\smallskip 
$

\textbf{Proof. }The statement ($a$) results from (\ref{Afa}) and Lemma 3.1($%
c $). In order to prove ($b$), note that for any $x\in C\cap \mathrm{int}\,%
\mathrm{dom}\,f$ and $y\in A^{f}x$ we have 
\begin{equation*}
D_{f}(z,x)-D_{f}(z,y)=D_{f}(y,x)-\left\langle \nabla f(x)-\nabla
f(y),z-y\right\rangle ,
\end{equation*}%
and $y=\nabla f^{\ast }(\nabla f(x)-$\thinspace $\xi )$ for some \thinspace $%
\xi \in Ax.$ Thus, $\nabla f(y)=\nabla f(x)-$\thinspace $\xi $ and%
\begin{equation}
D_{f}(z,x)-D_{f}(z,y)=D_{f}(y,x)+\left\langle 0^{\ast }-\,\xi
,z-y\right\rangle .  \label{vei}
\end{equation}%
If $z$ is a solution of Problem 1.1, then $z\in C,$ $0^{\ast }\in Az$ and $%
z=\nabla f^{\ast }(\nabla f(z)-0^{\ast }).$ By consequence,%
\begin{equation}
\left\langle 0^{\ast }-\,\xi ,z-y\right\rangle =\left\langle 0^{\ast }-\,\xi
,\nabla f^{\ast }(\nabla f(z)-0^{\ast })-\nabla f^{\ast }(\nabla f(x)-\,\xi
)\right\rangle .  \label{wei}
\end{equation}%
Since $A$ is $D_{f}$-coercive on $C$, it results that the right-hand side of
(\ref{wei}) is nonnegative (see (\ref{InvMon})) for any $x\in C\cap \mathrm{%
int}\,\mathrm{dom}\,f$. Hence, by (\ref{vei}) and (\ref{wei}), the
inequality in (\ref{tn}) results and it shows that $z\in \mathrm{Nexp}%
_{C}^{f}\,A^{f}.\hfill \square \medskip $

The class of operators which are $D_{f}$-coercive contains some meaningful
operators. Among them are all operators $A[T]:X\rightarrow 2^{X^{\ast }}$
given by%
\begin{equation}
A\left[ T\right] =\nabla f-\nabla f\circ T,  \label{AT}
\end{equation}%
where $T:X\rightarrow 2^{X}$ is a $D_{f}$-firm operator. In particular, $%
A[T] $ is $D_{f}$-coercive when $T=B_{f}$, where $B_{f}:X\rightarrow 2^{X}$
is the $D_{f}$\textit{-resolvent} (cf. \cite{BauBorCom-BregMon}) of a
monotone operator $B:X\rightarrow 2^{X^{\ast }}$, i.e.,%
\begin{equation}
B_{f}:=\left( \nabla f+B\right) ^{-1}\circ \nabla f.  \label{res}
\end{equation}%
According to \cite[Proposition 3.8]{BauBorCom-BregMon}, the operator $%
T=B_{f} $ satisfies the condition (\ref{DomT}). These facts are summarized
in the following lemma.\smallskip

\textbf{Lemma 3.4. }\textit{Let} $T:X\rightarrow 2^{X}$ \textit{be an
operator which satisfies }(\ref{DomT}).\textit{\ Then the following
statements are true:}

($a$)\textit{\ }$\mathrm{dom}\,A[T]=\mathrm{dom}\,T$\textit{\ and its
antiresolvent is }$A\left[ T\right] ^{f}=T$\textit{;}

($b$) \textit{The operator }$T$ \textit{is\ }$D_{f}$\textit{-firm if and
only if the operator }$A\left[ T\right] :X\rightarrow 2^{X^{\ast }}$ \textit{%
defined by} (\ref{AT}) \textit{is }$D_{f}$-\textit{coercive on its domain.}

($c$) \textit{If }$B:X\rightarrow 2^{X^{\ast }}$\textit{\ is a monotone
operator with} $\mathrm{dom}\,B\cap \mathrm{int}\,\mathrm{dom}\,f\neq
\varnothing $\textit{, then }$B_{f}$ is $D_{f}$\textit{-firm, single valued
on its domain, }%
\begin{equation}
A[B_{f}]=\nabla f-\nabla f\circ \left( \nabla f+B\right) ^{-1}\circ \nabla f,
\label{antires}
\end{equation}%
\textit{and }$A[B_{f}]$ \textit{is }$D_{f}$-\textit{coercive on its domain.}%
\smallskip

\textbf{Proof.} Statement ($a$) results from (\ref{Afa}) and (\ref{AT}). To
prove ($b$) observe that for any $x,y\in \mathrm{dom}\,T,$ for any
\thinspace $\xi \in A\left[ T\right] x$ and for any $\eta \in A\left[ T%
\right] y,$ we have 
\begin{equation}
\,\xi =\nabla f(x)-\nabla f(u)\text{\quad and\quad }\eta =\nabla f(y)-\nabla
f(v)  \label{gag}
\end{equation}%
for some $u\in Tx$ and for some $v\in Ty.$ Therefore,%
\begin{eqnarray}
&&\left\langle \,\xi -\eta ,\nabla f^{\ast }(\nabla f(x)-\,\xi )-\nabla
f^{\ast }(\nabla f(y)-\eta )\right\rangle  \label{blues} \\
&=&\left\langle \nabla f(x)-\nabla f(y),u-v\right\rangle -\left\langle
\nabla f(u)-\nabla f(v),u-v\right\rangle .  \notag
\end{eqnarray}%
If $T$ is $D_{f}$-firm, then the right hand side of (\ref{blues}) is
nonnegative and this implies that $A\left[ T\right] $ is $D_{f}$-coercive on 
$\mathrm{dom}\,T$. Conversely, suppose that $A\left[ T\right] $ is $D_{f}$%
-coercive on $\mathrm{dom}\,T$. If $u\in Tx$ and $v\in Ty,$ then the vectors
\thinspace $\xi $ and $\eta $ given by (\ref{gag}) satisfy (\ref{blues}) and
the left hand side of this equality is nonnegative. Hence, $T$ is $D_{f}$%
-firm. This proves ($b$). In order to prove ($c$) recall that, since $B$ is
monotone, its resolvent, $B_{f}$, is necessarily $D_{f}$-firm and single
valued on its domain (cf. \cite[Proposition 3.8]{BauBorCom-BregMon}) and, as
noted above, it satisfies (\ref{DomT}). Thus, according to ($b$), the
operator $A[B_{f}]$ is $D_{f}$-coercive on its domain.$\hfill \square
\medskip $

\subsection{Connection between the proximal-projection method and the
proximal point method}

Lemma 3.4 helps establishing a connection between the proximal-projection
method and the \textit{proximal point method (with Bregman distances)}. The
proximal point method we are referring to in this paper is the iterative
procedure which, in our setting, can be described by%
\begin{equation}
x^{0}\in \mathrm{dom}\,B_{f}\text{\quad and\quad }x^{k+1}=B_{f}(x^{k}),\text{
}\forall k\in \mathbb{N}\text{,}  \label{proxpt}
\end{equation}%
where $B:X\rightarrow 2^{X^{\ast }}$ is a monotone operator with $\mathrm{dom%
}\,B\cap \mathrm{int}\,\mathrm{dom}\,f\neq \varnothing .$ Its well
definedness is guaranteed when 
\begin{equation}
\varnothing \neq \mathrm{ran}\,B_{f}\subseteq \mathrm{dom}\,B_{f}.
\label{proxptconsist}
\end{equation}%
For ensuring that the inclusion in this condition holds it is sufficient to
make sure that $\mathrm{dom}\,B_{f}=X.$ This implicitly happens when one
considers the classical proximal point method (see \cite{RocProxPt}) where $%
X $ is a Hilbert space, $f=\textstyle%
\frac{1}{2}%
\left\Vert \cdot \right\Vert ^{2}$ and $B$ is presumed to be maximal
monotone. Alternative conditions which imply that $\mathrm{dom}\,B_{f}=X$
when $B$ is maximal monotone are presented in \cite{BurSch} in a more
general setting. In particular, those conditions hold if $X$ is a uniformly
convex and uniformly smooth Banach space, $B$ is maximal monotone and $f=%
\textstyle%
\frac{1}{2}%
\left\Vert \cdot \right\Vert ^{2}.$

Maximal monotonicity of $B$ is a commonly used condition for ensuring well
definedness of the proximal point method because, if $B$ is maximal monotone
and $f=\textstyle%
\frac{1}{2}%
\left\Vert \cdot \right\Vert ^{2},$ then $\nabla f+B$ and $\nabla f^{\ast }$
are surjective and, thus, $\nabla f^{\ast }\circ \left( \nabla f+B\right) $
is surjective too, that is, $\mathrm{dom}\,B_{f}=\mathrm{ran}\,\,\left[
\nabla f^{\ast }\circ \left( \nabla f+B\right) \right] =X.$ However, well
definedness of the proximal point method can be sometimes ensured for
operators $B$ which are monotone without being maximal monotone. In such
cases, it is interesting to know whether the proximal point method preserves
the convergence properties which make it so useful in applications requiring
finding zeros of maximal monotone operators. Here is an example of a
monotone operator which is not maximal and for which (\ref{proxptconsist})
holds (in spite of the fact that $\nabla f+B$ is not surjective). Take $X=%
\mathbb{R}$, $f=\textstyle%
\frac{1}{2}%
\left\vert \cdot \right\vert ^{2}$ and let $B:X\rightarrow 2^{X^{\ast }}$ be
given by $Bx=\left\{ 0\right\} $ if $x\leq 0$, and $Bx=\varnothing $ if $%
x>0. $ The operator $B$ is monotone, but it is not maximal monotone since
the operator defined by $B^{\prime }x=\left\{ 0\right\} $ for all $x\in X$
is a proper monotone extension of $B.$ Obviously, in this case $\nabla f$ is
the identity, $(\nabla f+B)x=\left\{ x\right\} $ if $x\leq 0,$ $(\nabla
f+B)x=\varnothing $ if $x>0$ and, therefore, $\nabla f+B$ is not surjective.
However, $B_{f}=\left( \nabla f+B\right) ^{-1}=\nabla f+B$ and, hence, $%
\mathrm{ran}\,B_{f}=\mathrm{dom}\,B_{f}=(-\infty ,0]$ showing that (\ref%
{proxptconsist}) is satisfied, that is, the proximal point algorithm is well
defined.

The next result establishes the connection between the proximal-projection
method and the proximal point method. It requires that $\mathrm{dom}\,B_{f}$
should be convex and closed. This necessarily happens if $\nabla f+B$ is
surjective and $\mathrm{dom}\,f=X.$ However, $\mathrm{dom}\,B_{f}$ may
happen to be convex and closed even if $\nabla f+B$ is not surjective as one
can see from the example above. Other instances in which $\mathrm{dom}%
\,B_{f} $ is convex and closed are described in the remarks preceding
Corollary 4.2 as well as in the body of that corollary.\smallskip

\textbf{Lemma 3.5. }\textit{Let\ }$B:X\rightarrow 2^{X^{\ast }}$\textit{\ be
a monotone operator such that }$\mathrm{dom}\,B_{f}$ \textit{is convex and
closed and suppose that }(\ref{proxptconsist}) \textit{is satisfied. Then
the proximal point method }(\ref{proxpt}) \textit{is exactly the
proximal-projection method applied to the }$D_{f}$\textit{-coercive operator 
}$A[B_{f}]$\textit{\ with }$C_{k}=X$ \textit{for all }$k\in \mathbb{N}$%
.\smallskip

\textbf{Proof. }Observe that, by (\ref{antires}), we have 
\begin{eqnarray*}
\mathrm{Proj}_{\mathrm{dom}\,B_{f}}^{f}\left( \nabla f-A[B_{f}]\right) &=&%
\mathrm{Proj}_{\mathrm{dom}\,B_{f}}^{f}\left[ \nabla f\circ \left( \nabla
f+B\right) ^{-1}\circ \nabla f\right] \\
&=&\mathrm{proj}_{\mathrm{dom}\,B_{f}}^{f}\left[ \left( \nabla f+B\right)
^{-1}\circ \nabla f\right] =B_{f},
\end{eqnarray*}%
where the last equality results holds because (\ref{proxptconsist}) is
satisfied. This shows that the proximal point method and the
proximal-projection method are overlapping when one takes $A=A[B_{f}]$ and $%
C_{k}=X$ for all $k\in \mathbb{N}$ in (\ref{alg}).$\hfill \square $

\section{Convergence Analysis of the Proximal-Projection Method}

In this section we present a convergence theorem for the proximal-projection
method in reflexive Banach spaces. Our convergence analysis is based on a
generalization of Lemma 5.7 in \cite{ButResSurvey} which, in turn, is a
generalization of a result known as Opial's Lemma \cite[Lemma 2]{opial}. All
over this section we assume that the function $f$ and the Banach space $X$
are as described in Section 1.

\subsection{A generalization of Opial's Lemma}

Opial's Lemma says that if $X$ is a Hilbert space and if $T:Y\rightarrow X$
is a nonexpansive mapping on the nonempty closed convex subset $Y$ of $X,$
then for any sequence $\left\{ z^{k}\right\} _{k\in \mathbb{N}}\subseteq Y$
which is weakly convergent and has $\lim_{k\rightarrow \infty }\left\Vert
Tz^{k}-z^{k}\right\Vert =0,$ the vector $z=w-\lim_{k\rightarrow \infty
}z^{k} $ is necessarily a fixed point of $T.$ In \cite[Lemma 5.7]%
{ButResSurvey} a similar result was shown to hold in Banach spaces which are
not necessarily Hilbert spaces. Namely, it was proved that the conclusion of
Opial's Lemma still holds for operators $T:X\rightarrow X$ which are
nonexpansive relative to $f$ (in the sense given to this term in \cite%
{ButReiZas2003}), i.e., such that%
\begin{equation}
D_{f}(Tx,Ty)\leq D_{f}(x,y),\quad \forall x,y\in Y,  \label{RelNonexp}
\end{equation}%
provided that $\mathrm{dom}\,f=\mathrm{dom}\,\nabla f=X$ and that $f$ it is
not only Legendre, but it is also totally convex (see \cite{ButIus-Book})
and bounded on bounded sets, while $T$ satisfies $\lim_{k\rightarrow \infty
}D_{f}(Tz^{k},z^{k})=0$.

Our current generalization of Opial's Lemma concerns set-valued operators $T$
satisfying a somehow less stringent nonexpansivity condition than (\ref%
{RelNonexp}) with respect to a function $f$ subjected to weaker requirements
than those involved in \cite[Lemma 5.7]{ButResSurvey}. In the sequels we use
the following notion which generalizes that of nonexpansive operator
relative to $f$.\smallskip

\textbf{Definition 4.1.} The operator $T:X\rightarrow 2^{X}$ is said to be $%
D_{f}$-\textit{nonexpansive }if it satisfies (\ref{DomT}) and for any $x\in 
\mathrm{dom}\,T$ there exists $u\in Tx$ such that 
\begin{equation}
(\forall y\in \mathrm{dom}\,T):\left[ v\in Ty\Rightarrow D_{f}(v,u)\leq
D_{f}(y,x)\right] .  \label{Dfnonexp}
\end{equation}

In that follows (see Theorem 4.1 below) we will be interested in operators
whose antiresolvents are simultaneously $D_{f}$-firm and $D_{f}$%
-nonexpansive. It should be noted that the notions of $D_{f}$-nonexpansivity
and $D_{f}$-firmness are not equivalent, although some operators may have
both properties. If $X$ is a Hilbert space provided with $f=\textstyle%
\frac{1}{2}%
\left\Vert \cdot \right\Vert ^{2},$ then it is obvious that any $D_{f}$-firm
operator (i.e., any firmly nonexpansive operator) is $D_{f}$-nonexpansive
(i.e., nonexpansive). Even in this context, the converse implication does
not generally hold. Take, for example, the case where the Hilbert space is $%
X=\mathbb{R}$ and $Ax=2x.$ Its antiresolvent is $A^{f}x=-x$ and it is $D_{f}$%
-nonexpansive without being $D_{f}$-firm. However, operators whose
antiresolvents are simultaneously $D_{f}$-nonexpansive and $D_{f}$-firm are
not specific to the setting of Hilbert spaces provided with $f=\textstyle%
\frac{1}{2}%
\left\Vert \cdot \right\Vert ^{2}$. For instance, if $X=\mathbb{R}$ and $%
f(x)=\textstyle%
\frac{1}{4}%
x^{4},$ then the antiresolvent $A^{f}$ of the operator $A=\alpha \nabla f$
with $\alpha \in (0,1)$ is $D_{f}$-nonexpansive and $D_{f}$-firm at the same
time. It is $D_{f}$-firm because, as shown in Section 3.2, $A$ is $D_{f}$%
-coercive and Lemma 3.2($c$) applies. $A^{f}$ is also $D_{f}$-nonexpansive
because%
\begin{equation*}
D_{f}(A^{f}y,A^{f}x)=(1-\alpha )^{4}D_{f}(y,x)\leq D_{f}(y,x),\text{ }%
\forall x,y\in X.
\end{equation*}%
One still may hope that (as happens in the particular situation noted above
when $X$ is a Hilbert space provided with $f=\textstyle%
\frac{1}{2}%
\left\Vert \cdot \right\Vert ^{2}$) a $D_{f}$-firm operator is always $D_{f}$%
-nonexpansive. The following example shows that this is not the
case.\smallskip

\textbf{Example 4.1. }\textit{A }$D_{f}$\textit{-firm} \textit{operator
which is not }$D_{f}$\textit{-nonexpansive. }Let $X=\mathbb{R}$ and let $f:%
\mathbb{R}\rightarrow \mathbb{R}$ be the Legendre function given by $%
f(x)=|x|^{3/2}.$ Take the continuous, single-valued operator $T:\mathbb{R}%
\rightarrow \mathbb{R}$ defined by

\begin{equation*}
T(x)=%
\begin{cases}
\frac{1}{4}, & \text{if }x<\frac{1}{16}, \\ 
\sqrt{x}, & \text{if }x\geq \frac{1}{16}.%
\end{cases}%
\end{equation*}%
We first show that $T$ is $D_{f}$-firm, that is, we verify that for every $%
x,y\in \mathbb{R}$ one has%
\begin{equation}
\lbrack f^{\prime }(T(x))-f^{\prime }(T(y))][T(x)-T(y)]\leq \lbrack
f^{\prime }(x)-f^{\prime }(y)][T(x)-T(y)].  \label{firm}
\end{equation}%
For symmetry reasons we can assume that $x>y$. The case $x,y<1/16$ being
trivial, we shall consider the following two cases.

\textit{Case 1.} $x,y\geq 1/16$. In this case (\ref{firm}) reduces to $[\sqrt%
[4]{x}-\sqrt[4]{y}][\sqrt{x}-\sqrt{y}]\leq (\sqrt{x}-\sqrt{y})^{2},$ which
is equivalent with the obviously true inequality $1\leq \sqrt[4]{x}+\sqrt[4]{%
y}$.

\textit{Case 2.} $x\geq 1/16$ and $y<1/16$. In this case we distinguish two
subcases:

($i$) $y\geq 0$. In this situation (\ref{firm}) can be re-written as $[\sqrt[%
4]{x}-\sqrt{1/4}][\sqrt{x}-1/4]\leq \lbrack \sqrt{x}-\sqrt{y}][\sqrt{x}%
-1/4], $ which, in turn, is equivalent with $\sqrt[4]{x}+\sqrt{y}\leq \sqrt{x%
}+1/2.$ This last inequality holds since for $x\geq 1/16$ and $y<1/16$ we
obviously have $\sqrt[4]{x}+\sqrt{y}\leq \sqrt[4]{x}+1/4$ and it is easy to
verify that 
\begin{equation}
\sqrt[4]{x}+1/4\leq \sqrt{x}+1/2,\text{ }\forall x\in \lbrack 0,\infty ).
\label{zumzum}
\end{equation}

(\textit{ii}) $y<0$. In this case, the inequality (\ref{firm}) is equivalent
to $[\sqrt[4]{x}-\sqrt{1/4}][\sqrt{x}-1/4]\leq \lbrack \sqrt{x}+\sqrt{-y}][%
\sqrt{x}-1/4],$ that is, to $\sqrt[4]{x}-1/2\leq \sqrt{x}+\sqrt{-y}.$ This
last inequality is true because $\sqrt{x}+\sqrt{-y}\geq \sqrt{x}$ and, by (%
\ref{zumzum}), we also have $\sqrt[4]{x}\leq \sqrt{x}+1/4<\sqrt{x}+1/2$ for
every $x>0.$

These show that the operator $T$ is $D_{f}$-firm. Now we verify that $T$ is
not $D_{f}$-nonexpansive, that is, that there exists two real numbers $x$
and $y$ such that%
\begin{equation}
f(T(y))-f(T(x))-f^{\prime }(T(x))(T(y)-T(x))>f(y)-f(x)-f^{\prime }(x)(y-x)
\label{nonex}
\end{equation}%
Taking $x=1/8$ and $y=1/16$, a simple computation shows that the left hand
side of (\ref{nonex}) equals $2^{13/2}[2^{3/4}+2^{1/2}-3]\simeq 8.6895,$
while the right hand side equals $2^{15/4}[2^{1/2}-1]\simeq 5.5730.\hfill
\square \smallskip $

Before proceeding with the presentation of our generalization of Opial's
Lemma several observations concerning its hypothesis are in order.\smallskip

\textbf{Remark 4.1. }($a$) The next result is a proper generalization of
Lemma 5.7 in \cite{ButResSurvey} which, in turn, is a proper generalization
of Opial's Lemma. To see this, observe that when the operator $%
T:X\rightarrow X$ is $D_{f}$-nonexpansive, it satisfies (\ref{equ2}) below.

($b$) The hypothesis of the next result implicitly require that the space $X$
should be reflexive. The fact is that a function $f$ which is lower
semicontinuous with $\mathrm{int}\,\mathrm{dom}\,f\neq \varnothing $ and
uniformly convex on bounded subsets of $\mathrm{int}\,\mathrm{dom}\,f$
exists on a Banach space $X$ only if $X$ is reflexive (cf. \cite[Corollary
4.3]{ButIusZal} in conjunction with \cite[Theorem 2.10($ii$)]{ButResSurvey}).%
$\smallskip $

With these facts in mind we now proceed with the presentation of the
generalization of Opial's Lemma.\smallskip

\textbf{Proposition 4.1.} \textit{Suppose that the function }$f$ \textit{is
uniformly convex on bounded subsets of }$\mathrm{int}\,\mathrm{dom}\,f.$%
\textit{\ Let }$T:X\rightarrow 2^{X}$\ \textit{be} \textit{an operator
satisfying} (\ref{DomT}) \textit{and suppose that} $\nabla f$\textit{\ is
bounded on bounded subsets of }$\mathrm{dom}\,T\cup \,\mathrm{ran}\,T$. 
\textit{If }$\{z^{k}\}_{k\in \mathbb{N}}\subseteq \mathrm{dom}\,T$\textit{\
is a sequence which converges weakly to a vector }$z\in \mathrm{dom}\,T$ 
\textit{and} \textit{if, for some sequence }$\left\{ u^{k}\right\} _{k\in 
\mathbb{N}}$\textit{\ satisfying} 
\begin{equation}
\left( \forall k\in \mathbb{N}:u^{k}\in Tz^{k}\right) \mathbb{\quad }\text{%
and\quad }\lim_{k\rightarrow \infty }D_{f}(u^{k},z^{k})=0,  \label{equ1}
\end{equation}%
\textit{there exists }$u\in T(z)$\textit{\ such that}%
\begin{equation}
\liminf_{k\rightarrow \infty }D_{f}(u^{k},u)\leq \liminf_{k\rightarrow
\infty }D_{f}(z^{k},z),  \label{equ2}
\end{equation}%
\textit{then the vector }$z$\textit{\ is a fixed point of }$T$\textit{.}%
\smallskip

\textbf{Proof}. For any $x\in \mathrm{int}\,\mathrm{dom}\,f$ one has%
\begin{equation*}
D_{f}(z^{k},x)-D_{f}(z^{k},z)=D_{f}(z,x)+\langle \nabla f(x)-\nabla
f(z),z-z^{k}\rangle .
\end{equation*}%
This implies that%
\begin{equation}
\liminf_{k\rightarrow \infty }D_{f}(z^{k},x)\geq
D_{f}(z,x)+\liminf_{k\rightarrow \infty }D_{f}(z^{k},z),  \label{nili}
\end{equation}%
because, since $\left\{ z^{k}\right\} _{k\in \mathbb{N}}$ converges weakly
to $z,$we have 
\begin{equation*}
\lim_{k\rightarrow \infty }\langle \nabla f(x)-\nabla f(z),z-z^{k}\rangle =0.
\end{equation*}%
Since the function $f$ is Legendre, it is strictly convex on $\mathrm{int}\,%
\mathrm{dom}\,f.$ This implies that $D_{f}(z,x)>0$ whenever $x\neq z$ (cf. 
\cite[Proposition 1.1.4]{ButIus-Book}) and, consequently, by (\ref{nili}) we
obtain%
\begin{equation}
x\neq z\Rightarrow \liminf_{k\rightarrow \infty
}D_{f}(z^{k},x)>\liminf_{k\rightarrow \infty }D_{f}(z^{k},z).  \label{eq:4}
\end{equation}%
We claim that%
\begin{equation}
\liminf_{k\rightarrow \infty }D_{f}(u^{k},u)=\liminf_{k\rightarrow \infty
}D_{f}(z^{k},u).  \label{eq:5}
\end{equation}%
To prove this claim, observe that%
\begin{equation}
D_{f}(u^{k},u)=D_{f}(z^{k},u)+[f(u^{k})-f(z^{k})]-\langle \nabla
f(u),u^{k}-z^{k}\rangle .  \label{eq:5'}
\end{equation}%
The sequence $\left\{ z^{k}\right\} _{k\in \mathbb{N}}$ is bounded as being
weakly convergent. Since the function $f$ is uniformly convex on bounded
subsets of $\mathrm{int}\,\mathrm{dom}\,f$, it is also sequentially
consistent (cf. \cite[Theorem 2.10]{ButResSurvey}). Therefore, by (\ref{equ1}%
), we deduce that 
\begin{equation}
\lim_{k\rightarrow \infty }\Vert u^{k}-z^{k}\Vert =0.  \label{eq:6}
\end{equation}%
Hence, $\left\{ u^{k}\right\} _{k\in \mathbb{N}}$ is bounded and 
\begin{equation}
\lim_{k\rightarrow \infty }\langle \nabla f(u),u^{k}-z^{k}\rangle =0.
\label{eq:7}
\end{equation}%
The convexity of $f$ on $\mathrm{int}\,\mathrm{dom}\,f$ implies%
\begin{equation}
\langle \nabla f(u^{k}),u^{k}-z^{k}\rangle \geq f(u^{k})-f(z^{k})\geq
\langle \nabla f(z^{k}),u^{k}-z^{k}\rangle ,\quad \forall k\in \mathbb{N}.
\label{eq:8}
\end{equation}%
By hypothesis, $\nabla f$ is bounded on bounded subsets of $\mathrm{dom}%
\,T\cup \,\mathrm{ran}\,T$. Therefore, the sequences $\{\nabla
f(z^{k})\}_{k\in \mathbb{N}}$ and $\{\nabla f(u^{k})\}_{k\in \mathbb{N}}$
are bounded. Thus, by (\ref{eq:6}) and (\ref{eq:8}), we deduce that%
\begin{equation*}
\lim_{k\rightarrow \infty }[f(u^{k})-f(z^{k})]=0.
\end{equation*}%
This, combined with (\ref{eq:5'}) and (\ref{eq:7}), implies (\ref{eq:5}) and
the claim above is proved.

Suppose by contradiction that $z\notin Tz$. Then the vector $u\in Tz$ whose
existence is guaranteed by hypothesis has $u\neq z$ and then, by (\ref{eq:4}%
), we deduce 
\begin{equation}
\liminf_{k\rightarrow \infty }D_{f}(z^{k},u)>\liminf_{k\rightarrow \infty
}D_{f}(z^{k},z).  \label{eq:9}
\end{equation}%
On the other hand, by (\ref{equ2}) and (\ref{eq:5}), we have that%
\begin{equation}
\liminf_{k\rightarrow \infty }D_{f}(z^{k},z)\geq \liminf_{k\rightarrow
\infty }D_{f}(u^{k},u)=\liminf_{k\rightarrow \infty }D_{f}(z^{k},u),
\label{eq:10}
\end{equation}%
which contradicts (\ref{eq:9}). This completes the proof.$\hfill \square
\medskip $

\subsection{A convergence theorem for the proximal-projection algorithm}

At this stage we are in position to consider the question of convergence of
the procedure (\ref{alg}) towards solutions of Problem 1.1. For this
purpose, we recall the following:\smallskip

\textbf{Definition 4.2.} (Cf. \cite{mosco}) ($a$) The \textit{weak upper
limit of the sequence} $\left\{ E_{k}\right\} _{k\in \mathbb{N}}$ of subsets
of $X$ is the set denoted \textrm{w-}$\overline{\lim }_{k\rightarrow \infty
}E_{k}$ and consisting of all $x\in X$ such that there exists a subsequence $%
\left\{ E_{i_{k}}\right\} _{k\in \mathbb{N}}$ of $\left\{ E_{k}\right\}
_{k\in \mathbb{N}}$ and a sequence $\left\{ x^{k}\right\} _{k\in \mathbb{N}}$
in $X$ which converges weakly to $x$ and has the property that $x^{k}\in
E_{i_{k}}$ for each $k\in \mathbb{N}$.

($b$) The operator $A:X\rightarrow 2^{X^{\ast }}$ is \textit{sequentially
weakly-strongly closed }if its graph is sequentially closed in $X\times
X^{\ast }$ provided with the $\left( weak\right) \times \left( strong\right) 
$-topology, that is, 
\begin{equation}
\left. 
\begin{array}{c}
\forall k\in \mathbb{N}:\text{ }\xi ^{k}\in Av^{k} \\ 
v^{k}\rightharpoonup v\text{ and }\xi ^{k}\rightarrow \xi%
\end{array}%
\right\} \Rightarrow \xi \in Av.  \label{wsc}
\end{equation}%
\smallskip

Before proceedings towards the main result of this paper the following
observations may be of use.\smallskip

\textbf{Remark 4.2. }($a$) The subset $E$ of $X$ may happen not to be convex
even if there exists a sequence of closed convex sets $\left\{ E_{k}\right\}
_{k\in \mathbb{N}}$ contained in $X$ such that \textrm{w-}$\overline{\lim }%
_{k\rightarrow \infty }E_{k}=E.$ Indeed, take $X=\mathbb{R}$ and 
\begin{equation*}
E_{k}:=\left\{ 
\begin{array}{cc}
\left[ -1-\textstyle%
\frac{1}{k+1}%
,-1+\textstyle%
\frac{1}{k+1}%
\right] & \text{if }k\text{ is even,} \\ 
&  \\ 
\left[ 1-\textstyle%
\frac{1}{k+1}%
,1+\textstyle%
\frac{1}{k+1}%
\right] & \text{if }k\text{ is odd.}%
\end{array}%
\right.
\end{equation*}%
It is easy to verify that the weak upper limit of this sequence of closed
convex sets is the nonconvex set $\left\{ -1,+1\right\} .$ This fact
explains why, in the next theorem, convexity of $C$ can not be derived from
the convexity of the sets $C_{k}$.

($b$) Any nonempty closed convex subset $E$ of $X$ is the weak upper limit
of a sequence of half spaces (corresponding to support hyperplanes)
containing it.

($c$) Among the operators which are sequentially weakly-strongly closed are
all the maximal monotone operators (see, for instance, \cite{PasSbu}).

($d$) An essential part of condition ($b$) of the theorem below is the
requirement that the gradient $\nabla f$ of the Legendre function $f$ should
be bounded on bounded subsets of\textit{\ }$\mathrm{int}\,\mathrm{dom}\,f.$
If the Legendre function $f$ has the property that $\nabla f$ is bounded on
bounded subsets of $\mathrm{int}\,\mathrm{dom}\,f,$ then $\mathrm{dom}\,f=X.$
Indeed, since $f$ is essentially smooth, it follows that $\mathrm{int}\,%
\mathrm{dom}\,f\neq \varnothing $ and for any sequence $\left\{
x^{k}\right\} _{k\in \mathbb{N}}$ contained in $\mathrm{int}\,\mathrm{dom}%
\,f $ and converging to a point of the boundary of $\mathrm{int}\,\mathrm{dom%
}\,f $ has the property that $\lim_{k\rightarrow \infty }\left\Vert \nabla
f(x^{k})\right\Vert _{\ast }=\infty $ (cf. \cite[Theorem 5.6]%
{BauBor-LegendreFcts}). Now, suppose that $\left\{ x^{k}\right\} _{k\in 
\mathbb{N}}$ is a convergent sequence contained in $\mathrm{int}\,\mathrm{dom%
}\,f$ and denote by $x$ its limit. Then the sequence $\left\{ \nabla f\left(
x^{k}\right) \right\} _{k\in \mathbb{N}}$ is bounded because the sequence $%
\left\{ x^{k}\right\} _{k\in \mathbb{N}}$ is bounded and $\nabla f$ is
bounded on bounded subsets of $\mathrm{int}\,\mathrm{dom}\,f.$ We claim that 
$x\in \mathrm{int}\,\mathrm{dom}\,f.$ Assume by contradiction that $x\notin 
\mathrm{int}\,\mathrm{dom}\,f.$ Then $x$ belongs to the boundary of $\mathrm{%
int}\,\mathrm{dom}\,f.$ Hence, $\lim_{k\rightarrow \infty }\left\Vert \nabla
f(x^{k})\right\Vert _{\ast }=\infty $ and this contradicts the boundedness
of $\left\{ \nabla f\left( x^{k}\right) \right\} _{k\in \mathbb{N}}.$ Since $%
\mathrm{int}\,\mathrm{dom}\,f$ contains the limit of any convergent sequence
of vectors contained in it, it follows that $\mathrm{int}\,\mathrm{dom}\,f$
is, simultaneously, a closed and open set. The space $X,$ being a Banach
space, it is necessarily arcways connected and, thus, a connected space (cf. 
\cite[Theorem 10.3.2]{czaszar}). Consequently, $X$ is the only nonempty
subset of $X$ which is open and closed at the same time (cf. \cite[Theorem
10.1.8]{czaszar}), that is, $\mathrm{int}\,\mathrm{dom}\,f=X.$\smallskip

The following theorem establishes the basic convergence properties of the
proximal-projection method. It should be observed that, in view of Remark
4.2($d$), the hypothesis of point ($b$) of the theorem implicitly requires
that $\mathrm{dom}\,f=X.$ At point ($ii$) of the theorem sequential
weak-weak continuity of $\nabla f$ is mentioned as a sufficient condition
for weak convergence (as opposed to subsequential convergence) of the
proximal-projection method. This condition is obviously satisfied whenever
the space $X$ has finite dimension. It is also satisfied if $X$ is a Hilbert
space and $f=\textstyle%
\frac{1}{2}%
\left\Vert \cdot \right\Vert ^{2}\ $as well as in some nonhilbertian Banach
spaces like $\ell ^{p}$ provided with $f=\textstyle%
\frac{1}{p}%
\left\Vert \cdot \right\Vert ^{p}$ for any $p\in (1,+\infty )$ -- see \cite[%
Proposition 8.2]{bro}.\smallskip

\textbf{Theorem 4.1. }\textit{Suppose that the function }$f$\textit{\ is
uniformly convex on bounded subsets of }$\mathrm{int}\,\mathrm{dom}\,f,$ $%
\nabla f^{\ast }$\textit{\ is bounded on bounded subsets of }$\nabla f(%
\mathrm{dom}\,A)$ \textit{and that, in addition to} (\ref{consist1}) \textit{%
and Assumption 1.1, the subsets} $C_{k}$\textit{\ of }$X$\textit{\ satisfy}%
\begin{equation}
C=\mathrm{w}\text{\textrm{-}}\overline{\lim }_{k\rightarrow \infty }C_{k}.
\label{ass3.9}
\end{equation}%
\smallskip \textit{If Problem 1.1\ has at least one solution, if the
operator }$A$\textit{\ is }$D_{f}$\textit{-coercive on the set }$%
Q:=\bigcup_{k\in \mathbb{N}}C_{k},$ \textit{and if} \textit{at least one of
the following two conditions is satisfied:\smallskip }

($a$) $\nabla f$ \textit{is uniformly continuous on bounded subsets of }$%
\mathrm{int}\,\mathrm{dom}\,f$\textit{\ and }$A$ \textit{is sequentially
weakly-strongly closed;}

($b$)\textbf{\ }$\nabla f$\textit{\ is bounded on bounded subsets of }$%
\mathrm{int}\,\mathrm{dom}\,f,$\textit{\ }$A^{f}$\textit{\ is}\textbf{\ }$%
D_{f}$\textit{-nonexpansive and }$C\subseteq \mathrm{dom}\,A$\textit{;}%
\smallskip

\noindent \textit{then} \textit{any sequence }$\{x^{k}\}_{k\in \mathbb{N}}$%
\textit{\ generated by the proximal-projection method }(\ref{alg})\textit{\
has the following properties:\smallskip }

($i$) \textit{It is bounded, has weak accumulation points and any such point
is a solution of} \textit{Problem 1.1;}

($ii$) \textit{If Problem 1.1\ has unique solution or if }$\nabla f$ \textit{%
is weakly-weakly} \textit{sequentially continuous, then the sequence }$%
\{x^{k}\}_{k\in \mathbb{N}}$\textit{\ converges weakly and its weak limit is
solution to Problem 1.1;}

($iii$) \textit{If the Banach space }$X$\textit{\ has finite dimension, then 
}$\{x^{k}\}_{k\in \mathbb{N}}$\textit{\ converges in norm to a solution of} 
\textit{Problem 1.1}.\smallskip

\textbf{Proof.} Let $z\in C$ be a solution of Problem 1.1. Then, clearly $%
z\in \mathrm{dom}\,A$ and, by (\ref{consist1}), we deduce that $z\in \mathrm{%
int}\,\mathrm{dom}\,f.$ For each $k\in \mathbb{N}$, let $\zeta ^{k}\in
Ax^{k} $ be such that%
\begin{equation}
x^{k+1}=\text{\textrm{Proj}}_{C_{k+1}\cap \mathrm{dom}\,A}^{f}(\nabla
f(x^{k})-\zeta ^{k}).  \label{mimu}
\end{equation}%
Denote by 
\begin{equation}
u^{k}:=\nabla f^{\ast }(\nabla f(x^{k})-\zeta ^{k}).  \label{uk}
\end{equation}%
Observe that 
\begin{equation}
u^{k}\in A^{f}x^{k}\quad \text{and\quad }x^{k+1}=\text{\textrm{proj}}%
_{C_{k+1}\cap \mathrm{dom}\,A}^{f}u^{k},\text{\quad }\forall k\in \mathbb{N}%
\text{.}  \label{procedure}
\end{equation}%
By hypothesis, the operator $A$ is $D_{f}$-coercive on the set $Q$ and $z$
is a solution of Problem 1.1. Note that, since $0^{\ast }\in Az$ and $z\in
C\subseteq Q$, Lemma 3.3 applies with $C$ replaced by $Q$. It implies that $%
z\in $\textrm{Nexp}$_{Q}^{f}\,A^{f}$. By Lemma 2.1 we have that 
\begin{equation}
x^{k}\in C_{k}\cap \mathrm{dom}\,A\cap \mathrm{int}\,\mathrm{dom}%
\,f\subseteq Q\cap \mathrm{dom}\,A\cap \mathrm{int}\,\mathrm{dom}\,f,\quad
\forall k\in \mathbb{N}.  \label{lili}
\end{equation}

First we prove the following:\smallskip

\textit{Claim 1: The sequence }$\left\{ x^{k}\right\} _{k\in \mathbb{N}}$%
\textit{\ is bounded.}\smallskip

In order to show this notice that, by applying Lemma 3.1($b$) to $z\in $%
\textrm{Nexp}$_{Q}^{f}\,A^{f}$ we deduce%
\begin{equation}
D_{f}(z,u^{k})+D_{f}(u^{k},x^{k})\leq D_{f}(z,x^{k}),\quad \forall k\in 
\mathbb{N}.  \label{3}
\end{equation}%
This implies%
\begin{equation}
D_{f}(z,u^{k})\leq D_{f}(z,x^{k}),\quad \forall k\in \mathbb{N}.  \label{3'}
\end{equation}%
By Assumption 1.1, we have that $z\in C\subseteq C_{k},$ for all $k\in 
\mathbb{N}.$ Thus, taking into account (\ref{Df}), (\ref{procedure}) and
Lemma 2.2 applied to $\varphi =\iota _{C_{k+1}\cap \,\mathrm{dom}\,A},$ we
obtain that%
\begin{equation}
D_{f}(z,x^{k+1})+D_{f}(x^{k+1},u^{k})\leq D_{f}(z,u^{k}),\quad \forall k\in 
\mathbb{N}\text{.}  \label{4}
\end{equation}%
Combining (\ref{4}) with (\ref{3'}) yields%
\begin{equation}
D_{f}(z,x^{k+1})\leq D_{f}(z,x^{k}),\quad \forall k\in \mathbb{N},  \label{5}
\end{equation}%
showing that the nonnegative sequence $\{D_{f}(z,x^{k})\}_{k\in \mathbb{N}}$
is nonincreasing and, therefore, bounded. Let $\beta $ be an upper bound of $%
\{D_{f}(z,x^{k})\}_{k\in \mathbb{N}}$. According to (\ref{Wf}), (\ref{Df})
and (\ref{5}), we deduce that%
\begin{equation*}
f^{\ast }(\nabla f(x^{k}))-\left\langle \nabla f(x^{k}),z\right\rangle
+f(z)=W_{f}(\nabla f(x^{k}),z)=D_{f}(z,x^{k})\leq \beta ,\quad \forall k\in 
\mathbb{N}\text{.}
\end{equation*}%
This implies that the sequence $\{\nabla f(x^{k})\}_{k\in \mathbb{N}}$ is
contained in the sublevel set\linebreak \textrm{lev}$_{\leq }^{\psi }\left(
\beta -f(z)\right) $ of the function $\psi :=$ $f^{\ast }-\left\langle \cdot
,z\right\rangle .$ Since $z\in \mathrm{int}\,\mathrm{dom}\,f=\mathrm{int}\,%
\mathrm{dom}\,(f^{\ast })^{\ast },$ and since the function $f^{\ast }$ is
proper and lower semicontinuous, application of the Moreau-Rockafellar
Theorem (\cite[Theorem 7(A)]{RocLevelSets} or \cite[Fact 3.1]%
{BauBorCom-Essential}) shows that $f^{\ast }-\left\langle \cdot
,z\right\rangle $ is coercive. Consequently, all sublevel sets of $\psi $
are bounded. Hence, the sequence $\{\nabla f(x^{k})\}_{k\in \mathbb{N}}$ is
bounded. By hypothesis, $\nabla f^{\ast }$ is bounded on bounded subsets of $%
\nabla f(\mathrm{dom}\,A)$ and, according to (\ref{alg}), $\left\{
x^{k}\right\} _{k\in \mathbb{N}}$ is contained in $\nabla f(\mathrm{dom}\,A)$%
. Hence, the sequence $x^{k}=\nabla f^{\ast }(\nabla f(x^{k})),$ $k\in 
\mathbb{N}$, is bounded. This proves Claim 1.

Now we are going to prove the following:\smallskip

\textit{Claim 2: The sequence }$\left\{ x^{k}\right\} _{k\in \mathbb{N}}$%
\textit{\ has weak accumulation points and any such point is a solution of
Problem 1.1}.\smallskip

The space $X$ being reflexive, there exists a weakly convergent subsequence $%
\{x^{i_{k}}\}_{k\in \mathbb{N}}$ of $\{x^{k}\}_{k\in \mathbb{N}}.$ Let ${%
\bar{x}=\,}\mathrm{w}$-$\lim_{k\rightarrow \infty }x^{i_{k}}.$ In order to
show that ${\bar{x}}\in C$, denote $y^{k}=x^{i_{k}}$ for each $k\in \mathbb{N%
}$. According to (\ref{lili}), we have that $y^{k}\in C_{i_{k}},$ for every $%
k\in \mathbb{N}.$ By hypothesis \textrm{w}-$\overline{\lim }_{k\rightarrow
\infty }C_{k}=C$ and this implies that ${\bar{x}}\in C$ (see Definition
4.2). It remains to prove that $0^{\ast }\in A{\bar{x}.}$ To this end,
observe that, according to (\ref{Df}), (\ref{4}) and (\ref{3'}) we have%
\begin{equation}
0\leq D_{f}(z,u^{k})-D_{f}(z,x^{k+1})\leq
D_{f}(z,x^{k})-D_{f}(z,x^{k+1}),\quad \forall k\in \mathbb{N}.  \label{6}
\end{equation}%
As noted above, the sequence $\{D_{f}(z,x^{k})\}_{k\in \mathbb{N}}$ is
nonincreasing and nonnegative and, therefore, it converges. By (\ref{6}),
this implies that the sequence $\{D(z,u^{k})\}_{k\in \mathbb{N}}$ converges
and has the same limit as $\{D_{f}(z,x^{k})\}_{k\in \mathbb{N}}$. By (\ref{3}%
) we also have that%
\begin{equation}
D_{f}(u^{k},x^{k})\leq D_{f}(z,x^{k})-D_{f}(z,u^{k}),\quad \forall k\in 
\mathbb{N},  \label{7}
\end{equation}%
and, thus,%
\begin{equation}
\lim_{k\rightarrow \infty }D_{f}(u^{k},x^{k})=0.  \label{8}
\end{equation}%
Since $f$ is uniformly convex on bounded subsets of $\mathrm{int}\,\mathrm{%
dom}\,f,$ it results that it is sequentially consistent too (cf. \cite[%
Theorem 2.10]{ButResSurvey}). Therefore, the equality (\ref{8}) implies that%
\begin{equation}
\lim_{k\rightarrow \infty }\left\Vert u^{k}-x^{k}\right\Vert =0.  \label{lm}
\end{equation}%
Now, we distinguish two possible situations. First, suppose that condition ($%
a$) is satisfied. Note that, according to (\ref{uk}), we have%
\begin{equation*}
\zeta ^{k}=\nabla f(x^{k})-\nabla f(u^{k}),\quad \forall k\in \mathbb{N}.
\end{equation*}%
Since $\nabla f$ is uniformly continuous on bounded subsets of its domain,
we deduce by (\ref{lm}) that $\lim_{k\rightarrow \infty }\zeta ^{k}=0^{\ast
}.$ The operator $A$ being sequentially weakly-strongly closed, this and the
fact that $\left\{ x^{i_{k}}\right\} _{k\in \mathbb{N}}$ converges weakly to 
$\bar{x}$ imply that $0^{\ast }\in A\bar{x}.$ Hence, $\bar{x}$ is a solution
of Problem 1.1 when condition ($a$) is satisfied.

Alternatively, suppose that condition ($b$) is satisfied. Recall that, in
this case, we necessarily have $\mathrm{dom}\,f=X$ (cf. Remark 4.2($d$)). By
hypothesis ($b$), we have that $C\subseteq \mathrm{dom}\,A.$ By Assumption
1.1, we have that 
\begin{equation*}
(\nabla f-A)(C)\subseteq (\nabla f-A)(C_{0})\subseteq \mathrm{int}\,\mathrm{%
dom}\,f^{\ast },
\end{equation*}%
and, as shown above, $\bar{x}\in C.$ Hence, 
\begin{equation*}
\varnothing \neq (\nabla f-A)(\bar{x})\subseteq \mathrm{int}\,\mathrm{dom}%
\,f^{\ast }
\end{equation*}
which implies that $\bar{x}\in \mathrm{dom}\,A^{f}.$ From (\ref{Dfnonexp})
written with $x^{i_{k}}$ instead of $y$ and $u^{i_{k}}$ instead of $v$, we
obtain that there exists ${\bar{u}}\in A^{f}{\bar{x}}$, such that%
\begin{equation}
D_{f}(u^{i_{k}},{\bar{u}})\leq D_{f}(x^{i_{k}},{\bar{x}}),\quad \forall k\in 
\mathbb{N}.  \label{9}
\end{equation}%
This implies%
\begin{equation}
\underset{k\rightarrow \infty }{\lim \inf }D_{f}(u^{i_{k}},{\bar{u}})\leq 
\underset{k\rightarrow \infty }{\lim \inf }D_{f}(x^{i_{k}},{\bar{x}}).
\label{kuk}
\end{equation}%
Proposition 4.1, (\ref{8}) and (\ref{kuk}) imply that ${\bar{x}}$ is a fixed
point of $A^{f}$, that is, $0^{\ast }\in A{\bar{x}.}$ Hence, ${\bar{x}}$ is
a solution of Problem 1.1. This completes the proof of ($i$).

The fact that if the Problem\ 1.1 has unique solution then the sequence $%
\left\{ x^{k}\right\} _{k\in \mathbb{N}}$ converges weakly to that solution
is an immediate consequence of ($i$). Assume that the function $f$ has
sequentially weakly-weakly continuous gradient. We show next that, in this
case, the sequence $\left\{ x^{k}\right\} _{k\in \mathbb{N}}$ can not have
more than one weak accumulation point. Suppose by contradiction that this is
not the case and that $x^{\prime }$ and $x^{\prime \prime }$ are two
different weak accumulation points of $\left\{ x^{k}\right\} _{k\in \mathbb{N%
}}.$ Let $\left\{ x^{i_{k}}\right\} _{k\in \mathbb{N}}$ and $\left\{
x^{j_{k}}\right\} _{k\in \mathbb{N}}$ be subsequences of $\left\{
x^{k}\right\} _{k\in \mathbb{N}}$ converging weakly to $x^{\prime }$ and $%
x^{\prime \prime }$, respectively. By ($i$) combined with Lemma 3.3($b$) it
results that $\left\{ x^{\prime },x^{\prime \prime }\right\} \subseteq 
\mathcal{S}_{f}(A,C)=$\textrm{Nexp}$_{C}^{f}(A).$ Hence, the inequality (\ref%
{3'}) still holds for any $z\in \left\{ x^{\prime },x^{\prime \prime
}\right\} .$ It implies that the sequences $\left\{ D_{f}(x^{\prime
},x^{k})\right\} _{k\in \mathbb{N}}$ and $\left\{ D_{f}(x^{\prime \prime
},x^{k})\right\} _{k\in \mathbb{N}}$ are convergent. Let $a$ and $b$ be
their respective limits. For any $k\in \mathbb{N}$ we have%
\begin{equation*}
D_{f}(x^{\prime },x^{k})-D_{f}(x^{\prime \prime },x^{k})=D_{f}(x^{\prime
},x^{\prime \prime })+\left\langle \nabla f(x^{\prime \prime })-\nabla
f(x^{k}),x^{\prime }-x^{\prime \prime }\right\rangle
\end{equation*}%
because of (\ref{Df}). Replacing in this equation $x^{k}$ by $x^{j_{k}}$ and
letting $k\rightarrow \infty $ we deduce that%
\begin{equation}
a-b=D_{f}(x^{\prime },x^{\prime \prime }),  \label{sisi}
\end{equation}%
because $\nabla f$ is sequentially weakly-weakly continuous. A similar
reasoning with $x^{\prime }$ and $x^{\prime \prime }$ interchanged shows that%
\begin{equation*}
b-a=D_{f}(x^{\prime \prime },x^{\prime }).
\end{equation*}%
Adding this equality with (\ref{sisi}) we obtain that $D_{f}(x^{\prime
},x^{\prime \prime })=0.$ This can not happen unless $x^{\prime }=x^{\prime
\prime }$ because the function $f$ is strictly convex on $C\cap \mathrm{int}%
\,\mathrm{dom}\,f$ as being Legendre. Thus, we reached a contradiction and
this completes the proof of ($ii$). It is clear that ($iii$) follows from ($%
i $) and ($ii$) since the gradient of any convex function in a finite
dimensional space is continuous on the interior of its domain (see, for
instance, \cite[Proposition 2.8]{PhelpsBook}). This completes the proof of
the theorem.$\hfill \square $

\subsection{Consequences of Theorem 4.1}

If $X$ is a Hilbert space provided with the function $f=\textstyle%
\frac{1}{2}%
\left\Vert \cdot \right\Vert ^{2},$ then Theorem 4.1 has a somewhat simpler
form and even strong convergence of the sequence generated by the
proximal-projection method can be sometimes ensured. Note that in this case
the operator $A:X\rightarrow 2^{X}$ is $D_{f}$-coercive if and only if it is 
\textit{firmly nonexpansive} in the sense that%
\begin{equation}
\left. 
\begin{array}{c}
x,y\in X \\ 
\xi \in Ax,\text{ }\zeta \in Ay%
\end{array}%
\right\} \Rightarrow \left\langle \xi -\zeta ,x-y\right\rangle \geq
\left\Vert \xi -\zeta \right\Vert ^{2}.  \label{fnonexp}
\end{equation}%
Clearly, if $A$ has this property, then the operator $A^{f}$ (which is
exactly $I-A$) is nonexpansive and, thus, $D_{f}$-nonexpansive. Since in
this situation \textrm{Proj}$_{C_{k}}^{f}$ is exactly the metric projection
operator \textrm{Proj}$_{C_{k}},$ we obtain the following result:\smallskip

\textbf{Corollary 4.1. }\textit{Let}\textbf{\ }$X$ \textit{be a Hilbert
space.} \textit{Suppose that }$A:X\rightarrow 2^{X}$\textit{\ is a firmly
nonexpansive operator (i.e., it satisfies} (\ref{fnonexp})\textit{)}. 
\textit{If }$C$ \textit{is a nonempty, closed and convex subset of} $X$%
\textit{,} \textit{if }$\left\{ C_{k}\right\} _{k\in \mathbb{N}}$\textit{\
is a sequence of subsets of }$X$\textit{\ satisfying} (\ref{ass3.9}) \textit{%
and such} \textit{that} $C_{k}\cap \mathrm{dom}\,A$ \textit{is convex and
closed and contains }$C$ \textit{for each }$k\in \mathbb{N}$\textit{, and if}
\textit{Problem 1.1\ has at least one solution, then the sequence }$\left\{
x^{k}\right\} _{k\in \mathbb{N}}$\textit{\ generated according to the rule}%
\begin{equation}
x^{0}\in C_{0}\cap \mathrm{dom}\,A\quad \text{and}\quad x^{k+1}\in \text{%
\textrm{Proj}}_{C_{k+1}\cap \mathrm{dom}\,A}(x^{k}-Ax^{k}),  \label{HilbPSM}
\end{equation}%
\textit{converges weakly to a solution of} \textit{Problem 1.1}.\textit{\ If 
}$\mathrm{int}\,\mathcal{S}_{f}(A,C)\neq \varnothing ,$\textit{\ then }$%
\left\{ x^{k}\right\} _{k\in \mathbb{N}}$\textit{\ converges strongly.}%
\smallskip

\textbf{Proof. }As noted above, the operator $A$ satisfying (\ref{fnonexp})
is $D_{f}$-coercive and $D_{f}$-nonexpansive. Applying Theorem 4.1($b$) with 
$f=\textstyle%
\frac{1}{2}%
\left\Vert \cdot \right\Vert ^{2}$ which has $\nabla f=I$ (and, hence, has $%
\nabla f$ sequentially weakly-weakly continuous) and taking into account
that $A^{f}=I-A$, one deduces that the sequence $\left\{ x^{k}\right\}
_{k\in \mathbb{N}}$ converges weakly to a vector in $\mathcal{S}_{f}(A,C).$
The set $\mathcal{S}_{f}(A,C)=\mathrm{Nexp}_{C}^{f}\,A^{f}$ is convex and
closed (cf. Lemma 3.1($a$)). In the current circumstances, the inequality (%
\ref{5}) still holds for all $z\in \mathcal{S}_{f}(A,C).$ It is equivalent
to the condition 
\begin{equation*}
\left\Vert z-x^{k+1}\right\Vert \leq \left\Vert z-x^{k}\right\Vert ,\text{%
\quad }\forall z\in \mathcal{S}_{f}(A,C),\text{ }\forall k\in \mathbb{N}%
\text{.}
\end{equation*}%
Therefore, one can apply Theorem 4.5.10 in \cite{BorZhu} and this result
implies that the sequence $\left\{ x^{k}\right\} _{k\in \mathbb{N}}$
converges strongly when $\mathrm{int}\,\mathcal{S}_{f}(A,C)\neq \varnothing $%
.$\hfill \square $\medskip

It was pointed out in Subsection 3.3 that there is a strong connection
between the proximal-projection method (\ref{alg}) and the proximal point
method (\ref{proxpt}) -- see Lemma 3.5. As far as we know, convergence of
the proximal point method in reflexive Banach spaces was established for
maximal monotone operators only. We use the connection between the proximal
point method and the proximal-projection method in order to obtain
convergence of the proximal point method for operators which are monotone
with sequentially weakly-weakly closed graphs (but are not necessarily
maximal monotone). Clearly, in spaces with finite dimension any monotone
operator with closed graph and, in general, monotone operators with closed
convex graphs have this property. The other requirement of the next
corollary that $\mathrm{dom}\,B_{f}$ should be convex is necessarily
satisfied if $\nabla f^{\ast }$ and $\nabla f+B$ are surjective because, in
this case, $\,\mathrm{dom}\,B_{f}=\,\mathrm{ran}\,\nabla f^{\ast }\circ
(\nabla f+B)=X.$ This condition is sufficient without being necessary as the
example preceding Lemma 3.5 shows. It can be easily verified that this also
happens whenever $\func{Graph}B$ is convex and $\nabla f$ is linear. Since
the corollary is based on Theorem 4.1($a$), the remarks preceding Theorem
4.2 concerning the implications of the hypothesis on the domains of $f$ and $%
f^{\ast }$ still apply here.\smallskip

\textbf{Corollary 4.2. }\textit{Suppose that the following conditions are
satisfied:\smallskip }

($a$)\textit{\ }$f$\textit{\ is uniformly convex on bounded subsets of} $%
\mathrm{int}\,\mathrm{dom}\,f$\textit{;}

($b$) $\nabla f$ \textit{is uniformly continuous on bounded subsets of }$%
\mathrm{int}\,\mathrm{dom}\,f$ \textit{as well as sequentially weakly to
weak continuous;}

($c$) $\nabla f^{\ast }$ \textit{is bounded on bounded subsets of} $\mathrm{%
int}\,\mathrm{dom}\,f^{\ast }$.\textit{\smallskip }

\noindent \textit{If }$B:X\rightarrow 2^{X^{\ast }}$\textit{\ is a monotone
operator with sequentially weakly-closed graph in }$X\times X^{\ast },$ 
\textit{satisfying }(\ref{proxptconsist}) \textit{and such that} $\mathrm{dom%
}\,B_{f}$ \textit{is convex}, \textit{if }$B$ \textit{has at least one zero
in }$\mathrm{int}\,\mathrm{dom}\,f$\textit{, and if either\smallskip }

($d$) $\mathrm{dom}\,B_{f}$ \textit{is closed in} $X$,

\noindent\textit{or}

($e$) $\nabla f$ \textit{is bounded on bounded subsets of} $\mathrm{int}\,%
\mathrm{dom}\,f,\smallskip $

\noindent\textit{then the sequences generated by the proximal point method }(%
\ref{proxpt}) \textit{converge weakly to zeros of the operator} $B$%
.\smallskip

\textbf{Proof. }We start by observing that, due to the boundedness on
bounded subsets of its domain of $\nabla f^{\ast },$ we have that $\mathrm{%
dom}\,f^{\ast }=X^{\ast }$ - see Remark 4.2($d$). We first prove that the
conclusion holds when $\mathrm{dom}\,B_{f}$ is closed in $X.$ Subsequently
we will show that, if $\nabla f$ is bounded on bounded subsets of $\mathrm{%
int}\,\mathrm{dom}\,f,$ then $\mathrm{dom}\,B_{f}$ is necessarily closed in $%
X$ and, thus, the conclusion is true in this case too.

So, assume that $\mathrm{dom}\,B_{f}$ is closed in $X.$ By Lemma 3.5, the
proximal point method is identical to the proximal-projection method applied
to the operator $A[B_{f}]$ given by (\ref{antires}). Therefore, for proving
the corollary in this case, it is sufficient to show that $A[B_{f}]$
satisfies the requirements of Theorem 4.1($a$). In order to do that it is
sufficient to ensure that the operator $A[B_{f}]$ is $D_{f}$-coercive on its
domain and has sequentially weakly-strongly closed graph. Observe that,
according to Lemma 3.4, the operator $A[B_{f}]$ is $D_{f}$-coercive on its
domain. It remains to show that $A[B_{f}]$ is sequentially weakly-strongly
closed. Let $\left\{ y^{k}\right\} _{k\in \mathbb{N}}$ be a weakly
convergent sequence contained in $\mathrm{dom}\,A[B_{f}]$ and denote $\xi
^{k}=A[B_{f}]y^{k}.$ Suppose that $\left\{ \xi ^{k}\right\} _{k\in \mathbb{N}%
}$ converges strongly in $X^{\ast }$ to some vector $\xi .$ Let $y=$w-$%
\lim_{k\rightarrow \infty }y^{k}.$ By hypothesis ($b$), the sequence $%
\left\{ \nabla f(y^{k})\right\} _{k\in \mathbb{N}}$ converges weakly in $%
X^{\ast }$ to $\nabla f(y)$.\ Thus, the sequence%
\begin{equation}
\nabla f\left( y^{k}\right) -\xi ^{k}=\left[ \nabla f\circ \left( \nabla
f+B\right) ^{-1}\circ \nabla f\right] (y^{k})  \label{plus}
\end{equation}%
converges weakly in $X^{\ast }$ to $\nabla f(y)-\xi .\ $Denote $%
u^{k}:=\nabla f^{\ast }\left( \nabla f\left( y^{k}\right) -\xi ^{k}\right) $
and observe that, by (\ref{plus}), we have that 
\begin{equation}
u^{k}=\left[ \left( \nabla f+B\right) ^{-1}\circ \nabla f\right]
(y^{k})=B_{f}y^{k},\text{ }\forall k\in \mathbb{N}\text{.}  \label{gal}
\end{equation}%
According to hypothesis ($c$), the sequence $\left\{ u^{k}\right\} _{k\in 
\mathbb{N}}$ is bounded because the sequence $\left\{ \nabla f\left(
y^{k}\right) -\xi ^{k}\right\} _{k\in \mathbb{N}}$ is bounded (as shown
above this sequence is weakly convergent). Let $\left\{ u^{i_{k}}\right\}
_{k\in \mathbb{N}}$ be a weakly convergent subsequence of $\left\{
u^{k}\right\} _{k\in \mathbb{N}}$ and let $u$ be the weak limit of this
subsequence. By (\ref{gal}) we deduce that $\nabla f(y^{i_{k}})\in \left(
\nabla f+B\right) u^{i_{k}}$ for all $k\in \mathbb{N}$, and thus we obtain%
\begin{equation}
\nabla f(y^{i_{k}})-\nabla f(u^{i_{k}})\in Bu^{i_{k}},\text{ }\forall k\in 
\mathbb{N}\text{.}  \label{baal}
\end{equation}%
By hypothesis ($b$), we have that 
\begin{equation}
\text{w-}\lim_{k\rightarrow \infty }\left[ \nabla f(y^{i_{k}})-\nabla
f(u^{i_{k}})\right] =\nabla f(y)-\nabla f(u).  \label{bal}
\end{equation}%
Since $\func{Graph}B$ is sequentially weakly-weakly closed and $\left\{
u^{i_{k}}\right\} _{k\in \mathbb{N}}$ converges weakly to $u,$ the relations
(\ref{baal}) and (\ref{bal}) imply that $\nabla f(y)-\nabla f(u)\in Bu,$
i.e., $\nabla f(y)\in \nabla f(u)+Bu.$ Consequently, we have that 
\begin{equation}
u=\left( \nabla f+B\right) ^{-1}\left( \nabla f(y)\right) =B_{f}y,
\label{plusplus}
\end{equation}%
because the operator $B_{f}$ is single valued (cf. Lemma 3.4). On the other
hand, by (\ref{plus}), (\ref{gal}), (\ref{bal}) and (\ref{plusplus}), we
have that%
\begin{equation*}
\xi =\nabla f(y)-\text{w-}\lim_{k\rightarrow \infty }\nabla
f(u^{i_{k}})=\nabla f(y)-\nabla f(u)=\nabla f(y)-\left( \nabla f\circ
B_{f}\right) y,
\end{equation*}%
showing that $(y,\xi )\in \func{Graph}A[B_{f}].$ Hence, $A[B_{f}]$ is
sequentially weakly-strongly closed and the proof, in this case, is complete.

Now, assume that $\nabla f$ is bounded on bounded subsets of $\mathrm{int}\,%
\mathrm{dom}\,f.$ Then, by Remark 4.2($d$), $\mathrm{dom}\,f=X$. We are
going to show that, in this case, the set $\mathrm{dom}\,B_{f}$ is closed.
As shown above, if $\mathrm{dom}\,B_{f}$ is closed, then the conclusion
holds. In order to prove that $\mathrm{dom}\,B_{f}$ is closed, let $\left\{
z^{k}\right\} _{k\in \mathbb{N}}$ be a sequence contained in $\mathrm{dom}%
\,B_{f}$ and converging in $X$ to some vector $\bar{z}.$ Denote $%
w^{k}=B_{f}z^{k}.$ We claim that the sequence $\left\{ w^{k}\right\} _{k\in 
\mathbb{N}}$ is bounded. To show that, note that, since $A[B_{f}]$ is $D_{f}$%
-coercive (cf. Lemma 3.4($c$)), all solutions of Problem 1.1 are in the set $%
\mathrm{Nexp}_{\,\mathrm{dom}\,A[B_{f}]}^{f}\left( A[B_{f}]\right) ^{f}$
(cf. Lemma 3.3). Taking into account that, by Lemma 3.4($a,b$), $%
B_{f}=\left( A[B_{f}]\right) ^{f}$ and $\mathrm{dom}\,B_{f}=\,\mathrm{dom}%
\,A[B_{f}],$ we deduce that all solutions of Problem 1.1 are contained in $%
\mathrm{Nexp}_{\mathrm{dom}\,B_{f}}^{f}\,B_{f}$. Let $z$ be such a solution.
Then%
\begin{equation*}
D_{f}(z,w^{k})+D_{f}(w^{k},z^{k})\leq D_{f}(z,z^{k}),\text{ }\forall k\in 
\mathbb{N}\text{.}
\end{equation*}%
According to the definition of the modulus of total convexity of $f$ on the
bounded set $\left\{ z^{k}\right\} _{k\in \mathbb{N}},$ denoted $\nu
_{f}(\left\{ z^{k}\right\} _{k\in \mathbb{N}},\cdot ),$ -- see \cite%
{ButResSurvey} -- we obtain 
\begin{equation}
0\leq \nu _{f}\left( \left\{ z^{k}\right\} _{k\in \mathbb{N}};\left\Vert
w^{k}-z^{k}\right\Vert \right) \leq D_{f}(w^{k},z^{k})\leq D_{f}(z,z^{k}),%
\text{ }\forall k\in \mathbb{N}\text{,}  \label{starlet}
\end{equation}%
Since $\nabla f$ is bounded on bounded subsets of $X,$ the function $f$ is
also bounded on bounded subsets of $X.$ Consequently, taking into account (%
\ref{Df}), we deduce that the sequence $\left\{ D_{f}(z,z^{k})\right\}
_{k\in \mathbb{N}}$ is bounded. Let $M$ be an upper bound of this sequence.
Suppose by contradiction that the sequence $\left\{ w^{k}\right\} _{k\in 
\mathbb{N}}$ contains a subsequence $\left\{ w^{j_{k}}\right\} _{k\in 
\mathbb{N}}$ such that $\lim_{k\rightarrow \infty }\left\Vert
w^{j_{k}}\right\Vert =\infty .$ Then there exists a positive integer $k_{0}$
such that for all integers $k\geq k_{0}$ we have $\left\Vert
w^{j_{k}}-z^{j_{k}}\right\Vert \geq 1.$ By \cite[Proposition 2.1($ii$)]%
{ButResSurvey} and (\ref{starlet}), we deduce that for any $k\geq k_{0}$ we
have%
\begin{equation}
\left\Vert w^{j_{k}}-z^{j_{k}}\right\Vert \nu _{f}\left( \left\{
z^{k}\right\} _{k\in \mathbb{N}};1\right) \leq \nu _{f}\left( \left\{
z^{k}\right\} _{k\in \mathbb{N}};\left\Vert w^{j_{k}}-z^{j_{k}}\right\Vert
\right) \leq M,\text{ }\forall k\in \mathbb{N}\text{.}  \label{toto}
\end{equation}%
The function $f$ is, by hypothesis, uniformly convex on bounded subsets of $%
\mathrm{int}\,\mathrm{dom}\,f$ and, consequently, it is also totally convex
on bounded subsets of $\mathrm{int}\,\mathrm{dom}\,f$ -- cf. \cite[Theorem
2.10]{ButResSurvey}. Therefore, $\nu _{f}\left( \left\{ z^{k}\right\} _{k\in 
\mathbb{N}};1\right) >0.$ Taking this into account together with the fact
that $\left\{ z^{j_{k}}\right\} _{k\in \mathbb{N}}$ is bounded (as being
convergent) and letting $k\rightarrow \infty $ in (\ref{toto}), we reach a
contradiction. Hence, the sequence $\left\{ w^{k}\right\} _{k\in \mathbb{N}}$
is bounded. Let $\left\{ w^{s_{k}}\right\} _{k\in \mathbb{N}}$ be a weakly
convergent subsequence of $\left\{ w^{k}\right\} _{k\in \mathbb{N}}$ and let 
$\bar{w}$ be the weak limit of this subsequence. According to (\ref{res}),
we have that%
\begin{equation}
\nabla f(z^{k})-\nabla f(w^{k})\in Bw^{k},\text{ }\forall k\in \mathbb{N}%
\text{.}  \label{moto}
\end{equation}%
Since $\nabla f$ is sequentially weakly-weakly continuous, and since $B$ has
sequentially weakly-weakly closed graph, the relation (\ref{moto}), written
with $s_{k}$ instead of $k,$ implies that $\nabla f(\bar{z})-\nabla f(\bar{w}%
)\in B\bar{w}.$ This shows that $\bar{w}=B_{f}\bar{z},$ that is, $\bar{z}\in 
\mathrm{dom}\,B_{f}$. Hence, $\mathrm{dom}\,B_{f}$ is closed and the proof
of the corollary is complete.$\hfill \square \medskip $

Another result which follows from Theorem 4.1 concerns a method of
regularizing and solving classical variational inequalities in the form (\ref%
{vi}). Since the problem of solving (\ref{vi}) may be ill-posed (in the
sense that it may not have solutions or it may have multiple solutions) and,
therefore, many algorithms for approximating solutions may not converge, or
may converge only subsequentially, to solutions of the problem, one
"regularizes" the original problem by solving an auxiliary problem which has
unique solution and whose solution is in the vicinity of the solution set of
the original problem, provided that the later is not empty. A regularization
technique, which originates in the works of Tikhonov \cite{tikhonov} and
Browder \cite{browder}, \cite{browder1}, consists of replacing the original
variational inequality (\ref{vi}) by the regularized variational inequality%
\begin{equation}
\text{Find }x\in C\cap \mathrm{int}\,\mathrm{dom}\,f\text{ such that}
\label{pertVI}
\end{equation}%
\begin{equation*}
\exists \,\xi \in Bx:\left[ \left\langle \xi +\alpha \nabla
f(x),y-x\right\rangle \geq 0,\text{ }\forall y\in C\cap \mathrm{dom}\,f%
\right] ,
\end{equation*}%
for some real number $\alpha >0.$ If $B$ is a monotone operator, then $%
B+\alpha \nabla f$ is strictly monotone and, therefore, the variational
inequality (\ref{pertVI}) can not have more then one solution. Moreover, in
many practically interesting situations, the variational inequality (\ref%
{pertVI}) has solution even if the original variational inequality (\ref{vi}%
) has not and, if $\alpha $ is sufficiently small, then the solution of (\ref%
{pertVI}) is close to the solution set of the unperturbed variational
inequality (\ref{vi}) whenever the later has solutions. This is, for
instance, the case (cf. \cite[Theorem 3.2]{AlbButRya2001}) when the Banach
space $X$ is simultaneously uniformly convex and uniformly smooth and
endowed with the Legendre function $f:=\textstyle%
\frac{1}{p}%
\left\Vert \cdot \right\Vert ^{p}$ for some $p>1$ and $B$ is maximal
monotone. Theorem 4.1 allows us to prove the next corollary which extends
the applicability of this regularization technique to reflexive Banach
spaces which are not necessarily uniformly convex and uniformly smooth and
to produce a weakly convergent algorithm for solving (\ref{pertVI}) in this,
more general setting, even if $B$ is not maximal monotone. This is of
interest because closedness of the solution of (\ref{pertVI}) with small $%
\alpha >0$ to the (presumed nonempty) solution set of the original
variational inequality (\ref{vi}) can be guaranteed even if $B$ is not
maximal monotone (cf. \cite[Theorem 2.1]{AlbButRya2005}).\smallskip

\textbf{Corollary 4.3. }\textit{Let }$B:X\rightarrow 2^{X^{\ast }}$\textit{\
be a monotone operator and let }$C$ \textit{be a nonempty, convex and closed
subset of} $\mathrm{dom}\,B\cap \mathrm{int}\,\mathrm{dom}\,f$\textit{.} 
\textit{Suppose that }$f$\textit{\ is uniformly convex on bounded subsets of 
}$\mathrm{int}\,\mathrm{dom}\,f,$ $\nabla f^{\ast }$\textit{\ is bounded on
bounded subsets of }$\mathrm{int}\,\mathrm{dom}\,f^{\ast }$ \textit{and
that, for some real number }$\alpha >0,$\textit{\ we have that}%
\begin{equation}
\varnothing \neq ((1-\alpha )\nabla f-B)(C)\subseteq \mathrm{int}\,\mathrm{%
dom}\,f^{\ast },  \label{Cconsist}
\end{equation}%
\textit{and} \textit{the operator }\textrm{Proj}$_{C}^{f}\left[ (1-\alpha
)\nabla f-B)\right] $\textit{\ is }$D_{f}$\textit{-firm.} \textit{If one of
the following conditions is satisfied:\smallskip }

($a$) $X$ \textit{has finite dimension, }$\mathrm{dom}\,f=X,$ $\nabla f$ 
\textit{is uniformly continuous on bounded subsets of} $X,$ \textit{and} $B$ 
\textit{has closed graph and is bounded on bounded subsets of its domain;}

($b$) $\nabla f$ \textit{is bounded on bounded subsets of} $\mathrm{int}\,%
\mathrm{dom}\,f,$ \textit{the operator}\linebreak \textrm{Proj}$_{C}^{f}%
\left[ (1-\alpha )\nabla f-B)\right] $\textit{\ is} $D_{f}$\textit{%
-nonexpansive;\smallskip }

\noindent \textit{then the iterative procedure defined by}%
\begin{equation}
x^{0}\in C\ \text{and}\ x^{k+1}\in \mathrm{Proj}_{C}^{f}\left[ (1-\alpha
)\nabla f(x^{k})-Bx^{k}\right] ,\ \forall k\in \mathbb{N},
\label{procedureVI}
\end{equation}%
\textit{is well defined and converges weakly to the necessarily unique
solution of the variational inequality} (\ref{pertVI})\textit{, provided
that such a solution exists.\smallskip }

\textbf{Proof. }Well definedness of the procedure results from (\ref%
{Cconsist}). The variational inequality (\ref{pertVI}) can not have more
than one solution since the operator $B^{\prime }:=B+\alpha \nabla f$ is
strictly monotone. According to Lemma 2.4, finding a solution of (\ref%
{pertVI}) is equivalent to finding a zero for the operator $V:=V[B^{\prime
};C;f]$ defined by (\ref{fakeB}). Note that%
\begin{equation}
V=\nabla f-\nabla f\circ \mathrm{Proj}_{C}^{f}((1-\alpha )\nabla f-B),
\label{V}
\end{equation}%
and%
\begin{equation}
V^{f}=\mathrm{Proj}_{C}^{f}((1-\alpha )\nabla f-B)  \label{Vf}
\end{equation}%
and that, by (\ref{star2}) applied to $B^{\prime }$ instead of $B$, we have%
\begin{equation}
\mathrm{Proj}_{C}^{f}\left( \nabla f-V\right) =\mathrm{Proj}%
_{C}^{f}((1-\alpha )\nabla f-B).  \label{VV}
\end{equation}%
Therefore, we can equivalently re-write the procedure (\ref{procedureVI}) as%
\begin{equation}
x^{0}\in C\quad \text{and}\quad x^{k+1}\in \mathrm{Proj}_{C}^{f}(\nabla
f\left( x^{k}\right) -Vx^{k}),\quad \forall k\in \mathbb{N}\text{.}
\label{procV}
\end{equation}%
This is exactly (\ref{alg}) applied to $V$ instead of $A$ with the sequence
of sets $C_{k}=C$ for all $k\in \mathbb{N}$. Now, suppose that condition ($a$%
) of our corollary is satisfied. In this case, if we show that the graph of $%
V$ is closed in $X\times X$ and that $V$ is $D_{f}$-coercive, then Theorem
4.1($a$) applies and leads to the conclusion of the corollary. Also, Theorem
4.1($b$) implies that, if condition ($b$) of the corollary holds, then the
procedure (\ref{procV}) is weakly convergent to the unique solution of (\ref%
{pertVI}), provided that $V$ is $D_{f}$-coercive. $D_{f}$-coercivity of $V$
results in both cases from Lemma 3.2 combined with (\ref{Vf}) and with our
hypothesis that $V^{f}=$\textrm{Proj}$_{C}^{f}\left[ (1-\alpha )\nabla f-B)%
\right] $ is $D_{f}$-firm. So, it remains to prove that, under assumption ($%
a $) of the corollary, the graph of $V$ is closed. To this end, let $\left\{
y^{k}\right\} _{k\in \mathbb{N}}$ be a sequence in $\mathrm{dom}\,V$ and
assume that this sequence converges to $y\in X.$ Let $\left\{ \xi
^{k}\right\} _{k\in \mathbb{N}}$ be the sequence%
\begin{equation*}
\xi ^{k}=\nabla f(y^{k})-\nabla f\circ \mathrm{Proj}_{C}^{f}((1-\alpha
)\nabla f(y^{k})-\zeta ^{k}),
\end{equation*}%
where $\zeta ^{k}\in By^{k}$ for all $k\in \mathbb{N}$. Suppose that $%
\lim_{k\rightarrow \infty }\xi ^{k}=\xi .$ Then, by Lemma 2.1, we have%
\begin{equation*}
\nabla f(y^{k})-\xi ^{k}=\left[ \nabla f\circ \left( \nabla f+N_{C}\right)
^{-1}\right] ((1-\alpha )\nabla f(y^{k})-\zeta ^{k}).
\end{equation*}%
This implies that%
\begin{equation}
(1-\alpha )\nabla f(y^{k})-\zeta ^{k}\in \left( \nabla f+N_{C}\right) \left[
\nabla f^{\ast }\left( \nabla f(y^{k})-\xi ^{k}\right) \right] ,\text{ }%
\forall k\in \mathbb{N}\text{.}  \label{hau}
\end{equation}%
Since $B$ is bounded on the bounded set $\left\{ y^{k}\right\} _{k\in 
\mathbb{N}},$ it follows that the sequence $\left\{ \zeta ^{k}\right\}
_{k\in \mathbb{N}}$ is bounded. Let $\left\{ \zeta ^{i_{k}}\right\} _{k\in 
\mathbb{N}}$ be a convergent subsequence of $\left\{ \zeta ^{k}\right\}
_{k\in \mathbb{N}}$ and let $\zeta $ be its limit. Since $\nabla f$ and $%
\nabla f^{\ast }$ are continuous on their respective domains and the
normality operator $N_{C}$ is maximal monotone (and, hence, has closed
graph), (\ref{hau}) implies%
\begin{equation*}
(1-\alpha )\nabla f(y)-\zeta \in \left( \nabla f+N_{C}\right) \left[ \nabla
f^{\ast }\left( \nabla f(y)-\xi \right) \right] .
\end{equation*}%
Therefore, we have%
\begin{equation*}
\nabla f^{\ast }\left( \nabla f(y)-\xi \right) =\mathrm{Proj}_{C}^{f}\left[
(1-\alpha )\nabla f(y)-\zeta \right]
\end{equation*}%
showing that%
\begin{equation*}
\xi =\nabla f(y)-\nabla f\circ \mathrm{Proj}_{C}^{f}\left[ (1-\alpha )\nabla
f(y)-\zeta \right] \in Vy.
\end{equation*}%
This completes the proof.$\hfill \square \medskip $

The requirement made in Corollary 4.3 that the operator $V^{f}=\linebreak 
\mathrm{Proj}_{C}^{f}\left[ (1-\alpha )\nabla f-B\right] $ should be $D_{f}$%
-firm may seem unusual. Here are several examples which show that this
condition is quite often satisfied without excessively costly demands on the
operator $B$ or the function $f.$ The next example shows that Corollary 4.3($%
b$) is applicable to solve variational inequalities in the form (\ref{pertVI}%
) when $X$ is any Hilbert space, $f=\textstyle%
\frac{1}{2}%
\left\Vert \cdot \right\Vert ^{2},$ $C=\mathrm{dom}\,B=X$ and $B$ is a
monotone operator which is either contractive with some constant $\gamma >0$
or strongly monotone with some constant $\delta >0$ and even in more general
conditions (when (\ref{cutu}) holds for some $\alpha \in (0,\textstyle%
\frac{1}{2}%
)$ and $\beta >0$)$.$ Obviously, in this setting, (\ref{vi}) is exactly the
problem of finding a zero of $B.$ To follow the considerations in the
examples below one should first note that by replacing in (\ref{vi}) the
operator $B$ by $\beta B,$ where $\beta $ is a positive constant, one
obtains a variational inequality which is equivalent to the original one.$%
\smallskip $

\textbf{Example 4.2. }\textit{Let }$X$\textit{\ be a Hilbert space and let} $%
f=\textstyle%
\frac{1}{2}%
\left\Vert \cdot \right\Vert ^{2}.$ \textit{Suppose that} $C=\mathrm{dom}%
\,B=X.$ \textit{If }$B$\textit{\ is contractive with some constant }$\gamma
>0$\textit{\ or strongly monotone with some constant }$\delta >0,$\textit{\
then the operator} $V^{f}=\mathrm{Proj}_{C}^{f}\left[ (1-\alpha )\nabla
f-\beta B\right] $ \textit{is }$D_{f}$\textit{-firm and }$D_{f}$\textit{%
-nonexpansive whenever} $\alpha \in (0,\textstyle%
\frac{1}{2}%
)$ \textit{with} $\beta =\sqrt{\alpha (1-\alpha )}\gamma ^{-1}$ \textit{in
the contractive case and} $\beta =(1-2\alpha )\delta $ \textit{in the
strongly monotone case.}

In order to prove this observe that, in the current setting, $\nabla f=%
\mathrm{Proj}_{C}^{f}=I.$ Also, the $D_{f}$-firmness condition (\ref{Dfirm})
for $T=V^{f}$ is equivalent to (\ref{fnonexp}) and this is exactly%
\begin{equation*}
\left\langle \left[ (1-\alpha )x-\beta \xi \right] -\left[ (1-\alpha
)y-\beta \eta \right] ,x-y\right\rangle
\end{equation*}%
\begin{equation*}
\geq \left\Vert \left[ (1-\alpha )x-\beta \xi \right] -\left[ (1-\alpha
)y-\beta \eta \right] \right\Vert ^{2},
\end{equation*}%
which can be equivalently re-written as%
\begin{equation}
\alpha (1-\alpha )\left\Vert x-y\right\Vert ^{2}+(1-2\alpha )\beta
\left\langle \xi -\eta ,x-y\right\rangle \geq \beta ^{2}\left\Vert \xi -\eta
\right\Vert ^{2},  \label{cutu}
\end{equation}%
for all pairs $(x,\xi ),$ $(y,\eta )\in \func{Graph}B.$ Moreover, if $V^{f}$
is $D_{f}$-firm, then it is also $D_{f}$-nonexpansive. Due to the
monotonicity of $B$ the inequality (\ref{cutu}) holds whenever $\alpha \in
(0,\textstyle%
\frac{1}{2}%
)$ and 
\begin{equation}
\alpha (1-\alpha )\left\Vert x-y\right\Vert \geq \beta \left\Vert \xi -\eta
\right\Vert ,\text{ }\forall (x,\xi )\in \func{Graph}B,\forall (y,\eta )\in 
\func{Graph}B.  \label{newcontractive}
\end{equation}%
If $B$ is contractive with constant $\gamma ,$ then we have%
\begin{equation*}
\gamma \left\Vert x-y\right\Vert \geq \left\Vert \xi -\eta \right\Vert ,%
\text{ }\forall (x,\xi )\in \func{Graph}B,\forall (y,\eta )\in \func{Graph}B
\end{equation*}%
and multiplying this inequality by $\beta =\sqrt{\alpha (1-\alpha )}\gamma
^{-1}$ we deduce that (\ref{newcontractive}) holds. Similarly, note that (%
\ref{cutu}) is satisfied when%
\begin{equation}
(1-2\alpha )\left\langle \xi -\eta ,x-y\right\rangle \geq \beta \left\Vert
\xi -\eta \right\Vert ^{2},\text{ }\forall (x,\xi )\in \func{Graph}B,\forall
(y,\eta )\in \func{Graph}B.  \label{newStrongMon}
\end{equation}%
If $B$ is strongly monotone with constant $\delta ,$ then%
\begin{equation*}
\left\langle \xi -\eta ,x-y\right\rangle \geq \delta \left\Vert \xi -\eta
\right\Vert ^{2},\text{ }\forall (x,\xi )\in \func{Graph}B,\forall (y,\eta
)\in \func{Graph}B.
\end{equation*}%
Multiplying this inequality by $(1-2\alpha )$ we deduce that (\ref%
{newStrongMon}) holds in this case.$\hfill \square \medskip $

In situations in which $X$ is not a Hilbert space, or $X$ is a Hilbert space
but the monotone operator $B$ does not satisfy (\ref{cutu}) for some $\alpha
\in (0,\textstyle%
\frac{1}{2}%
)$ and $\beta >0$, the question of how to choose the "regularization
function" $f$ and the "regularization parameter" $\alpha $ in the perturbed
variational inequality (\ref{pertVI}) in order to force $D_{f}$-firmness
and/or $D_{f}$-nonexpansivity on $V^{f}=\mathrm{Proj}_{C}^{f}((1-\alpha
)\nabla f-B)$ is relevant. Here are examples of situations when $V^{f}$ is $%
D_{f}$-firm even if $X$ is not a Hilbert space.\smallskip

\textbf{Example 4.3. }\textit{If the monotone operator }$B:X\rightarrow
2^{X^{\ast }}$\textit{\ and the nonempty closed convex subset }$C$ \textit{of%
} $\mathrm{dom}\,B\cap \mathrm{int}\,\mathrm{dom}\,f$\textit{\ have the
property }(\ref{Cconsist}), then $V^{f}=\mathrm{Proj}_{C}^{f}((1-\alpha
)\nabla f-B)$ \textit{is }$D_{f}$\textit{-firm in any of the following
situations:}

($a$) $C=X,$ $\alpha \in (0,1)$ \textit{and the operator }$\alpha \nabla f+B$%
\textit{\ is }$D_{f}$\textit{-coercive on its domain;}

($b$) $X$\textit{\ is a Hilbert space, }$C=X,$\textit{\ }$f=\textstyle%
\frac{1}{2}%
\left\Vert \cdot \right\Vert ^{2}$\textit{, }$\alpha \in (0,\textstyle%
\frac{1}{2}%
]$\textit{\ and }$B$\textit{\ is contractive with constant }$\alpha
(1-\alpha )$\textit{.}

In order to show this observe that in case ($a$), according to Lemma 2.1, we
have $V^{f}=(\alpha \nabla f+B)^{f}$ and, by Lemma 3.2($c$), the conclusion
follows. In case ($b$) an easy calculation shows that the operator $\alpha
\nabla f+B$ is $D_{f}$-coercive and, therefore, the conclusion of case ($a$)
applies.$\hfill \square \smallskip $

Example 4.3($a$), in conjunction with Corollary 4.3, leads to an algorithm
for solving the variational inequality (\ref{pertVI}) whenever is possible
to find a Legendre function $f$ and a number $\alpha \in (0,1)$ such that $%
\alpha \nabla f+B$ is $D_{f}$-coercive on its domain. Example 4.3($b$)
covers the class of expansive operators in Hilbert spaces for which, as far
as we know, few methods of finding zeros are available.\smallskip

\textbf{Example 4.4. }\textit{If the monotone operator }$B:X\rightarrow
2^{X^{\ast }}$\textit{\ and the nonempty closed convex subset }$C$ \textit{of%
} $\mathrm{dom}\,B\cap \mathrm{int}\,\mathrm{dom}\,f$\textit{\ have the
property }(\ref{Cconsist}) \textit{and if}%
\begin{equation}
\left\langle \left[ \alpha \nabla f(x)+\xi \right] -\left[ \alpha \nabla
f(y)+\eta \right] ,\right.  \label{suzi}
\end{equation}%
\begin{equation*}
\left. \mathrm{Proj}_{C}^{f}\left[ (1-\alpha )\nabla f(x)-\xi \right] -%
\mathrm{Proj}_{C}^{f}\left[ (1-\alpha )\nabla f(y)-\eta \right]
\right\rangle \geq 0,
\end{equation*}%
\textit{for any }$(x,\xi )$\textit{\ and }$(y,\eta )$\textit{\ in} $\func{%
Graph}B$\textit{, then the operator} $V^{f}=\mathrm{Proj}_{C}^{f}\left[
(1-\alpha )\nabla f-B\right] $ \textit{is }$D_{f}$\textit{-firm. In
particular, this happens in any of the following situations:}

($a$) $B=\left( \textstyle%
\frac{1}{2}%
-\alpha \right) \nabla f$ \textit{and} $\alpha \in (0,\textstyle%
\frac{1}{2}%
)$\textit{;}

($b$) \textit{The operator }$P:X\times X^{\ast }\rightarrow 2^{X^{\ast
}\times X},$\textit{\ defined by} 
\begin{equation}
P(z,\zeta )=(0^{\ast },\mathrm{Proj}_{C}^{f}((1-\alpha )\nabla f(z)-\zeta )),
\label{uzi}
\end{equation}%
\textit{for some} $\alpha \in (0,1)$ \textit{is monotone when }$X\times
X^{\ast }$\textit{\ is provided with the norm }$\left\Vert (z,\zeta
)\right\Vert =\left( \left\Vert z\right\Vert ^{2}+\left\Vert \zeta
\right\Vert _{\ast }^{2}\right) ^{1/2}$\textit{\ and with the duality
pairing }$\left\langle (z,\zeta ),(\zeta ^{\prime },z^{\prime
})\right\rangle =\left\langle \zeta ^{\prime },z\right\rangle +\left\langle
\zeta ,z^{\prime }\right\rangle $\textit{\ (and, therefore, its dual is
isometric with }$X^{\ast }\times X$\textit{).}

Let $x,y\in \mathrm{dom}\,V^{f}$ and let $\xi \in Bx$ and $\eta \in By.$
Denote%
\begin{equation*}
x^{\prime }=\mathrm{Proj}_{C}^{f}((1-\alpha )\nabla f(x)-\xi )\text{ and }%
y^{\prime }=\mathrm{Proj}_{C}^{f}((1-\alpha )\nabla f(y)-\eta ).
\end{equation*}%
By Lemma 2.1 we have%
\begin{equation*}
(1-\alpha )\nabla f(x)-\xi -\nabla f(x^{\prime })\in N_{C}(x^{\prime })\text{
and }(1-\alpha )\nabla f(y)-\eta -\nabla f(y^{\prime })\in N_{C}(y^{\prime
}),
\end{equation*}%
which imply%
\begin{equation*}
\left\langle (1-\alpha )\nabla f(x)-\xi ,y^{\prime }-x^{\prime
}\right\rangle \leq \left\langle \nabla f(x^{\prime }),y^{\prime }-x^{\prime
}\right\rangle
\end{equation*}%
and, respectively,%
\begin{equation*}
\left\langle (1-\alpha )\nabla f(y)-\eta ,x^{\prime }-y^{\prime
}\right\rangle \leq \left\langle \nabla f(y^{\prime }),x^{\prime }-y^{\prime
}\right\rangle .
\end{equation*}%
Summing up the last two inequalities we obtain that%
\begin{equation}
\left\langle \left[ (1-\alpha )\nabla f(x)-\xi \right] -\left[ (1-\alpha
)\nabla f(y)-\eta \right] ,x^{\prime }-y^{\prime }\right\rangle  \label{luzi}
\end{equation}%
\begin{equation*}
\geq \left\langle \nabla f(x^{\prime })-\nabla f(y^{\prime }),x^{\prime
}-y^{\prime }\right\rangle .
\end{equation*}%
The $D_{f}$-firmness condition (\ref{Dfirm}) for the operator $T=V^{f}$ is
exactly%
\begin{equation*}
\left\langle \nabla f(x)-\nabla f(y),x^{\prime }-y^{\prime }\right\rangle
\geq \left\langle \nabla f(x^{\prime })-\nabla f(y^{\prime }),x^{\prime
}-y^{\prime }\right\rangle .
\end{equation*}%
According to (\ref{luzi}), this is satisfied when%
\begin{equation*}
\left\langle \nabla f(x)-\nabla f(y),x^{\prime }-y^{\prime }\right\rangle
\geq \left\langle \left[ (1-\alpha )\nabla f(x)-\xi \right] -\left[
(1-\alpha )\nabla f(y)-\eta \right] ,x^{\prime }-y^{\prime }\right\rangle
\end{equation*}%
and this last inequality is equivalent to (\ref{suzi}). Hence, if (\ref{suzi}%
) holds, then the operator $V^{f}$ is $D_{f}$-firm. In case ($a$) we have
that $\alpha \nabla f+B=(1-\alpha )\nabla f-B$ and using the monotonicity of 
$\mathrm{Proj}_{C}^{f}$ (cf. \cite[Theorem 4.6]{ButResSurvey}) one deduces
that (\ref{suzi}) holds. Suppose that we are in case ($d$) and the operator $%
P,$ given by (\ref{uzi}), is monotone. Then observe that%
\begin{equation*}
\left\langle \left[ \alpha \nabla f(x)+\xi \right] -\left[ \alpha \nabla
f(y)+\eta \right] ,\right.
\end{equation*}%
\begin{equation*}
\left. \mathrm{Proj}_{C}^{f}\left[ (1-\alpha )\nabla f(x)-\xi \right] -%
\mathrm{Proj}_{C}^{f}\left[ (1-\alpha )\nabla f(y)-\eta \right]
\right\rangle =
\end{equation*}%
\begin{equation*}
\left\langle (x,\alpha \nabla f(x)+\xi )-(y,\alpha \nabla f(y)+\eta
),P(x,\xi )-P(y,\eta )\right\rangle ,
\end{equation*}%
and that the last expression is nonnegative due to the monotonicity of $P$
(see \cite[Proposition 4.7]{ButResSurvey}). Hence, (\ref{suzi}) holds in
this case too.$\hfill \square \smallskip $

Note that problem (\ref{pertVI}) in which $B=\beta \nabla f$ for some $\beta
>0$ is equivalent to the problem of finding the minimizer of $f$ over $C.$
The facts observed in Example 4.4($a$), in conjunction with Corollary 4.3,
leads to a proximal-projection method of finding that minimizer, provided
that $f$ satisfies the other requirements there. Obviously, the
effectiveness of that method, as well as of the other methods discussed in
this work, depends on the possibility of computing proximal projections onto 
$C.$ Algorithms for computing proximal projections are presented in \cite%
{AlbBut}, \cite{BauCom-BregBestAp} and \cite{ButResSurvey}. Compared with
already classical projection methods (see \cite{polyakBook}) in which the
iterations are, usually, of the form $x^{k+1}=\mathrm{Proj}%
_{C}(x^{k}-\lambda _{k}\nabla f(x^{k}))$ with a converging to zero positive
step size $\lambda _{k},$ the proximal-projection method presents the
advantage of not requiring arbitrarily small step sizes which, in practical
applications, may force the procedure to became stationary long before the
iterates are close to the minimizer of $f$ (due to the computer
identification of $\lambda _{k}\nabla f(x^{k})$ with the null vector).

In general, it would be nice to have a Legendre function $f$ for which the
condition in Example 4.4($b$) is satisfied. Whether such a function exists
in a Banach space $X$ is an \textit{open question}. This question is
relevant because such a function $f$, if any, would be a "universal
regularizer" for variational inequalities in the form (\ref{vi}) in the
sense that it would be such that the regularized variational inequality (\ref%
{pertVI}) will be solvable by the proximal-projection method, no matter how
the monotone operator $B$ is, provided that (\ref{pertVI}) has solution.

\section{Convergence of the Proximal-Projection Method in Spaces of Finite
Dimension}

Theorem 4.1 and its corollaries ensure weak and, sometimes, strong
convergence of the proximal-projection method to solutions of the Problem
1.1 under conditions which, besides the $D_{f}$-coercivity of the operator $%
A,$ require sequential weak-strong closedness of the $\func{Graph}A$ or,
alternatively, $D_{f}$-nonexpansivity of $A^{f}.$ In this section we show
that, when the space $X$ has finite dimension, some of these requirements
can be dropped or weakened. This is possible due to the validity in spaces
of finite dimension of another generalization of Opial's Lemma which we
present below.

\subsection{Another variant of a generalized Opial's Lemma.}

The following result applies to operators $T:X\rightarrow 2^{X}$ which are
not necessarily $D_{f}$-nonexpansive, but satisfy condition (\ref{eqn2})
below which is more general than $D_{f}$-firmness (compare condition (\ref%
{eqn2}) with Definition 3.3). It is interesting to observe that, if $f$ is
uniformly convex on bounded subsets of $\mathrm{int}\,\mathrm{dom}\,f$ and
if $T$ has closed graph, then the conclusion of the next result holds even
if the hypothesis that $u$ satisfies (\ref{eqn2}) is removed. This happens
because the equality in (\ref{eqn1}) implies that the sequences $\left\{
z^{k}\right\} _{k\in \mathbb{N}}$ and $\left\{ u^{k}\right\} _{k\in \mathbb{N%
}}$ converge to the same limit $z$ and, then, closedness of the graph of $T$
guarantees that $z\in Tz.$\smallskip

\textbf{Proposition 5.1. }\textit{Suppose that the space }$X$\textit{\ has
finite dimension, }$f$\textit{\ is uniformly convex on bounded subsets of}$\,%
\mathrm{int}\,\mathrm{dom}\,f$\textit{\ and }$T:X\rightarrow 2^{X}$ \textit{%
is an operator satisfying condition} (\ref{DomT})\textit{. Let} $%
\{z^{k}\}_{k\in \mathbb{N}}$ \textit{be a sequence in} $\mathrm{dom}\,T$ 
\textit{converging to an element }$z\in \mathrm{dom}\,T$. \textit{If for} 
\textit{some sequence} $\{u^{k}\}_{k\in \mathbb{N}}$ \textit{satisfying}%
\begin{equation}
(\forall k\in \mathbb{N}:u^{k}\in Tz^{k})\quad \text{and}\quad
\lim_{k\rightarrow \infty }D_{f}(u^{k},z^{k})=0,  \label{eqn1}
\end{equation}%
\textit{there exists} $u\in Tz$\textit{\ such that}%
\begin{equation}
\liminf_{k\rightarrow \infty }\,\langle \nabla f(u^{k})-\nabla
f(u),u^{k}-u\rangle \leq 0,  \label{eqn2}
\end{equation}%
\textit{then the vector }$z$\textit{\ is a fixed point of }$T$\textit{%
.\smallskip }

\textbf{Proof.} Since the function $f$ is convex and differentiable on $%
\mathrm{int}\,\mathrm{dom}\,f,$ the gradient $\nabla f$ is continuous on $%
\mathrm{int}\,\mathrm{dom}\,f.$ This fact and the strict convexity of $f$ on 
$\mathrm{int}\,\mathrm{dom}\,f$ imply that%
\begin{equation}
\lim_{k\rightarrow \infty }\langle \nabla f(z^{k})-\nabla
f(x),z^{k}-x\rangle =\langle \nabla f(z)-\nabla f(x),z-x\rangle >0.
\label{eqn3}
\end{equation}%
whenever $x\in \left( \mathrm{int}\,\mathrm{dom}\,f\right) \backslash
\left\{ z\right\} .$ The function $f$ being uniformly convex on bounded
subsets of $\mathrm{int}\,\mathrm{dom}\,f,$ it is also sequentially
consistent (cf. \cite[Theorem 2.10]{ButResSurvey}). Therefore, the equality
in (\ref{eqn1}) implies that 
\begin{equation}
\lim_{k\rightarrow \infty }\left\Vert z^{k}-u^{k}\right\Vert =0.  \label{bon}
\end{equation}%
Consequently, the sequences $\{z^{k}\}_{k\in \mathbb{N}}$ and $%
\{u^{k}\}_{k\in \mathbb{N}}$ converge to the same limit $z.$ By condition (%
\ref{DomT}), the boundedness of $\{u^{k}\}_{k\in \mathbb{N}}$ and the
continuity of $\nabla f$ we deduce that 
\begin{equation}
\lim_{k\rightarrow \infty }\langle \nabla f(z^{k})-\nabla
f(z),z^{k}-z\rangle =0=\lim_{k\rightarrow \infty }\langle \nabla
f(z^{k})-\nabla f(z),u^{k}-u\rangle .  \label{eq4new}
\end{equation}%
Note that%
\begin{eqnarray}
\langle \nabla f(u^{k})-\nabla f(u),u^{k}-u\rangle &=&\langle \nabla
f(u^{k})-\nabla f(u),u^{k}-z^{k}\rangle  \label{lon} \\
&&+\langle \nabla f(u^{k})-\nabla f(u),z^{k}-u\rangle  \notag \\
&=&\langle \nabla f(u^{k})-\nabla f(u),u^{k}-z^{k}\rangle  \notag \\
&&+\langle \nabla f(z^{k})-\nabla f(u),z^{k}-u\rangle  \notag \\
&&+\langle \nabla f(u^{k})-\nabla f(z^{k}),z^{k}-u\rangle .  \notag
\end{eqnarray}%
Since the sequence $\{\nabla f(u^{k})\}_{k\in \mathbb{N}}$ is bounded, it
follows from (\ref{bon}) that the first term of the last sum in (\ref{lon})
converges to zero. The second term of the same sum is nonnegative because of
the monotonicity of $\nabla f.$ Taking the limit as $k\rightarrow \infty $
on both sides of (\ref{lon}), we obtain that%
\begin{equation}
\lim_{k\rightarrow \infty }\langle \nabla f(u^{k})-\nabla
f(u),u^{k}-u\rangle \geq \lim_{k\rightarrow \infty }\langle \nabla
f(z^{k})-\nabla f(u),z^{k}-u\rangle .  \label{eqn5}
\end{equation}

In order to conclude the proof, suppose by contradiction that $z\notin T(z)$%
. Then, $u\neq z$ and, therefore, by (\ref{eq4new}), (\ref{eqn2}), (\ref%
{eqn5}) and (\ref{eqn3}), respectively, we obtain 
\begin{eqnarray*}
0 &=&\lim_{k\rightarrow \infty }\langle \nabla f(z^{k})-\nabla
f(z),z^{k}-z\rangle =\lim_{k\rightarrow \infty }\langle \nabla
f(z^{k})-\nabla f(z),u^{k}-u\rangle \\
&\geq &\lim_{k\rightarrow \infty }\langle \nabla f(u^{k})-\nabla
f(u),u^{k}-u\rangle \geq \lim_{k\rightarrow \infty }\langle \nabla
f(z^{k})-\nabla f(u),z^{k}-u\rangle >0,
\end{eqnarray*}%
which is a contradiction.$\hfill \square $

\subsection{A convergence theorem for the proximal-projection method in
spaces of finite dimension.}

The following theorem shows that, in finite dimensional spaces, convergence
of the proximal-projection method to solutions of Problem 1.1 can be ensured
with lesser requirements on the operator $A$ in addition to the $D_{f}$%
-coercivity than those involved in Theorem 4.1 and its corollaries.\smallskip

\textbf{Theorem 5.1. }\textit{Suppose that the space }$X$\textit{\ has
finite dimension, }$f$\textit{\ is uniformly convex on bounded subsets of }$%
\mathrm{int}\,\mathrm{dom}\,f,$ $\nabla f^{\ast }$\textit{\ is bounded on
bounded subsets of }$\nabla f(\mathrm{dom}\,A)$ \textit{and that }(\ref%
{consist1})\textit{, Assumption 1.1 hold and}%
\begin{equation}
C\cap \mathrm{dom}\,A=\mathrm{w}\text{\textrm{-}}\overline{\lim }%
_{k\rightarrow \infty }\left( C_{k}\cap \mathrm{dom}\,A\right) .
\label{pimp}
\end{equation}%
\textit{If the Problem 1.1 has at least one solution, if the operator} $%
A:X\rightarrow 2^{X^{\ast }}$ \textit{is }$D_{f}$\textit{-coercive on }$%
Q=\dbigcup\limits_{k=0}^{\infty }C_{k}$\textit{\ and if }$C\cap \mathrm{dom}%
\,A$ \textit{is closed, then the sequences generated by} \textit{the
proximal-projection method} (\ref{alg}) \textit{are well defined and
converge to solutions of} \textit{the Problem 1.1}.\smallskip

\textbf{Proof. }Well definedness of the sequences generated by (\ref{alg})
follows from (\ref{consist1}) and\textit{\ }Assumption 1.1. Suppose that,
for each $k\in \mathbb{N}$, $\zeta ^{k}\ $and $u^{k}$ are as in (\ref{mimu})
and (\ref{uk}), respectively. Then, clearly, condition (\ref{procedure})
holds too. The operator $A$ being $D_{f}$-coercive on its domain, the
operator $A^{f}$ is $D_{f}$-firm (cf. Lemma 3.2). This means that%
\begin{equation}
\langle \nabla f(u^{k})-\nabla f(u),u^{k}-u\rangle \leq \left\langle \nabla
f(x^{k})-\nabla f(x),u^{k}-u\right\rangle ,  \label{good}
\end{equation}%
for any pair $(x,u)\in \func{Graph}A^{f}$ and for any $k\in \mathbb{N}.$
Now, repeating without change the arguments in the proof of Theorem 4.1 one
can see that Claim 1 proven there still holds in our setting and implies
that the sequence $\left\{ x^{k}\right\} _{k\in \mathbb{N}}$ is bounded. Let 
$\left\{ x^{i_{k}}\right\} _{k\in \mathbb{N}}$ be a convergent subsequence
of $\left\{ x^{k}\right\} _{k\in \mathbb{N}}$ and let $\bar{x}$ be its
limit. An argument identical to that made in the proof of Theorem 4.1 (Claim
2) for the same purpose shows that $\bar{x}\in C$ and (\ref{lm}) holds.
According to (\ref{pimp}), since $x^{i_{k}}\in C_{i_{k}}\cap \mathrm{dom}%
\,A, $ it results that $\bar{x}\in C\cap \mathrm{dom}\,A.$ Hence, $A\bar{x}%
\neq \varnothing .$ Writing (\ref{good}) for $i_{k}$ instead of $k$ and $%
\bar{x}$ instead of $x$, and for any $u\in A\bar{x},$ letting in the
resulting inequality $k\rightarrow \infty $\ and taking into account that $%
\nabla f$ is continuous on $\mathrm{int}\,\mathrm{dom}\,f,$ we deduce that%
\begin{equation}
\underset{k\rightarrow \infty }{\lim \inf }\langle \nabla
f(u^{i_{k}})-\nabla f(u),u^{i_{k}}-u\rangle \leq 0.  \label{verygood}
\end{equation}%
This shows that the sequence $\left\{ u^{i_{k}}\right\} _{k\in \mathbb{N}}$
satisfies (\ref{eqn2}) for $T=A^{f},$ $z=\bar{x}$ and any $u\in A^{f}\bar{x}%
. $ Also, by (\ref{lm}) and (\ref{verygood}), the sequences $\left\{
x^{i_{k}}\right\} _{k\in \mathbb{N}}$ and $\left\{ u^{i_{k}}\right\} _{k\in 
\mathbb{N}}$ satisfy (\ref{eqn1}). Hence, Proposition 5.1 applies to $%
T=A^{f},$ $z=\bar{x}$ and $u\in A^{f}\bar{x}.$ By consequence, we have that $%
\bar{x}$ is a fixed point of $A^{f}$ and, hence, a zero of $A.$ It remains
to prove that $\bar{x}$ is the only accumulation point of the sequence $%
\left\{ x^{k}\right\} _{k\in \mathbb{N}}.$ The proof in this respect
reproduces without modifications the arguments made in the proof of Theorem
4.1 in order to show that the sequence $\left\{ x^{k}\right\} _{k\in \mathbb{%
N}}$ has a single weak accumulation point when $\nabla f$ is sequentially
weakly-weakly continuous.$\hfill \square \medskip $

\subsection{Consequences of Theorem 5.1.}

Using Theorem 5.1 instead of Theorem 4.1 we can prove again Corollary 4.2
and Corollary 4.3 in a finite dimensional setting with different and less
demanding conditions on $A$. Here is the new version of Corollary
4.2.\smallskip

\textbf{Corollary 5.1. }\textit{Suppose that the space }$X$\textit{\ has
finite dimension, }$f$\textit{\ is uniformly convex on bounded subsets of }$%
\mathrm{int}\,\mathrm{dom}\,f$ \textit{and} $\nabla f^{\ast }$\textit{\ is
bounded on bounded subsets of }$\mathrm{int}\,\mathrm{dom}\,f^{\ast }.$ 
\textit{If }$B:X\rightarrow 2^{X^{\ast }}$\textit{\ is a monotone operator
satisfying }(\ref{proxptconsist}) \textit{and having at least one zero, and
if any of the following conditions holds}

($a$) $\mathrm{ran}\,\left( \nabla f+B\right) $ \textit{is closed in }$%
X^{\ast }$ \textit{and} $\nabla f^{\ast }\left( \mathrm{ran}\,\left( \nabla
f+B\right) \right) $ \textit{is convex;}

($b$) $\mathrm{ran}\,\left( \nabla f+B\right) =X^{\ast }$\textit{;}

\noindent\textit{then the sequences generated by the proximal point method }(%
\ref{proxpt}) \textit{converge to zeros of the operator} $B$.\smallskip

\textbf{Proof.} Recall that in this setting $\mathrm{dom}\,f^{\ast }=X^{\ast
}$ (cf. Remark 4.2($d$)). Observe that, according to Lemma 3.4, we have that%
\begin{equation*}
\mathrm{dom}\,A\left[ B_{f}\right] =\mathrm{dom}\,B_{f}=\nabla f^{\ast
}\left( \mathrm{ran}\,\left( \nabla f+B\right) \right) .
\end{equation*}%
Therefore, if condition ($a$) holds, the set $\mathrm{dom}\,A\left[ B_{f}%
\right] =\mathrm{dom}\,B_{f}$ is closed and convex. Since condition ($b$)
implies ($a$), this remains true when ($b$) holds. Again by Lemma 3.4, the
operator $A[B_{f}]$ is $D_{f}$-coercive on its domain. Consequently, the
operator $A:=A\left[ B_{f}\right] $ satisfies the requirements of Theorem
5.1 when $C=C_{k}=X$ for all $k\in \mathbb{N}$. Applying Theorem 5.1 to $%
A[B_{f}]$ and taking into account Lemma 3.5 the conclusion follows.$\hfill
\square \medskip $

Now we give another variant of Corollary 4.3($a$) in which the condition
that $A$ should have closed graph is replaced by less demanding
requirements.\smallskip

\textbf{Corollary 5.2. }\textit{Let }$B:X\rightarrow 2^{X^{\ast }}$\textit{\
be a monotone operator.} \textit{Suppose that the space }$X$\textit{\ has
finite dimension, }$f$\textit{\ is uniformly convex on bounded subsets of }$%
\mathrm{int}\,\mathrm{dom}\,f$ \textit{and} $\nabla f^{\ast }$\textit{\ is
bounded on bounded subsets of }$\mathrm{int}\,\mathrm{dom}\,f^{\ast }.$ 
\textit{If }$C$\textit{\ is a closed convex subset of }$\mathrm{dom}\,B\cap 
\mathrm{int}\,\mathrm{dom}\,f$ \textit{such that, for some real number }$%
\alpha >0$\textit{,}%
\begin{equation}
\varnothing \neq ((1-\alpha )\nabla f-B)(C)\subseteq \mathrm{int}\,\mathrm{%
dom}\,f^{\ast },  \label{Cconsistnew}
\end{equation}%
\textit{and the operator }\textrm{Proj}$_{C}^{f}\circ \left[ (1-\alpha
)\nabla f-B)\right] $\textit{\ is }$D_{f}$\textit{-firm,} \textit{then the
iterative procedure given by }(\ref{procedureVI})\textit{\ is well defined
and converges to the necessarily unique solution of the variational
inequality} (\ref{pertVI})\textit{, provided that such a solution
exists.\smallskip }

\textbf{Proof. }Since (\ref{Vf}) and (\ref{VV}) still hold, the operator $V$
given by (\ref{V}) is $D_{f}$-coercive on its domain (cf. Lemma 3.4). By (%
\ref{Cconsistnew}) and by the fact that $C\subseteq \mathrm{dom}\,B\cap 
\mathrm{int}\,\mathrm{dom}\,f,$ it results that $C\subseteq \mathrm{dom}\,V.$
Hence, Theorem 5.1 applies to the operator $A=V$ and the sets $C_{k}=C$ and
the conclusion follows.$\hfill \square $


\bigskip

\begin{thebibliography}{99}
\bibitem{Alber1993} Alber, Ya.I, Generalized projection operators in Banach
spaces: Properties and applications, in: "Functional Differential Equations"
edited by M.E. Draklin and E. Litsyn, Vol. 1, The research Institute of
Judea and Samaria, Kedumim-Ariel, 1993.

\bibitem{alberGenproj} Alber Ya.I.: Metric and generalized projection
operators in Banach spaces: properties and applications, in "Theory and
Applications of Nonlinear Operators of Accretive and Monotone Type", pp.
15--50, edited by A. G. Kartsatos, Lecture Notes in Pure and Appl. Math.,
178, \textit{Marcel Dekker}, New York, 1996.

\bibitem{alber3} Alber, Ya.I.: Generalized projections, decompositions, and
the Pythagorean-type theorem in Banach spaces, \textit{Appl. Math. Lett.} 
\textbf{11} (1998), 115-121.

\bibitem{alberStab} Alber, Ya.I.: Stability of the proximal projection
algorithm for nonsmooth convex optimization problems with perturbed
constraint sets, \textit{J. Nonlinear Convex Anal.} \textbf{4 }(2003), 1--14.

\bibitem{AlbBut} Alber, Ya.I. and Butnariu, D.: Convergence of Bregman
projection methods for solving consistent convex feasibility problems in
reflexive Banach spaces. \textit{J. Optim. Theory Appl.} \textbf{92} (1997),
33--61.

\bibitem{AlbButKas} Alber, Ya.I., Butnariu, D. and Kassay, G.: On the
convergence and stability of a regularization method for maximal monotone
inclusions and its applications to convex optimization, in: F. Giannessi and
G. Maugeri (eds.) "Variational Inequalities and Applications", pp. 89-132, 
\textit{Springer} \textit{Verlag, }New York 2005.

\bibitem{AlbButRya2001} Alber, Ya.I., Butnariu, D. and Ryazantseva, I.:
Regularization methods for ill-posed inclusions and variational inequalities
with domain perturbations, \textit{J. Convex Nonlin. Anal. }\textbf{2}
(2001), 53-79.

\bibitem{AlbButRya2004} Alber, Ya.I., Butnariu, D. and Ryazantseva, I.:
Regularization of monotone variational inequalities with Mosco
approximations of the constraint sets, \textit{Set-Valued Anal.} \textbf{13}
(2005), 265--290.

\bibitem{AlbButRya2005} Alber, Ya.I., Butnariu, D. and Ryazantseva, I.:
Regularization and resolution of monotone variational inequalities with
operators given by hypomonotone approximations, \textit{J. Nonlinear Convex
Anal.} \textbf{6} (2005), 23--53.

\bibitem{AlberGuerre2001} Alber, Ya.I. and Guerre-Delabriere, S.: On the
projection method for fixed point problems, \textit{Analysis} \textit{%
(Munich)} \textbf{21} (2001), 17-39.

\bibitem{alber-ius-sol} Alber, Ya.I., Iusem, A.N. and Solodov, M.V.:
Minimization of nonsmooth convex functionals in Banach spaces, \textit{J.
Convex Anal.} \textbf{4} (1997), 235-255.

\bibitem{alb-kar-lit} Alber, Y.I., Kartsatos, A.G. and Litsyn E.: Iterative
solutions of unstable variational inequalities on approximately given sets, 
\textit{Abstr. Appl. Anal.} \textbf{1} (1996), 45-54.

\bibitem{alb-nash} Alber, Ya.I. and Nashed, Z.M.: Iterative-projection
regularization of ill-posed variational inequalities, \textit{Analysis
(Munich)} \textbf{24} (2004), 19--39.

\bibitem{alber-rei} Alber, Ya.I. and Reich, S.: An iterative method for
solving a class of nonlinear operator equations in Banach spaces, \textit{%
Panamer. Math. J.} \textbf{4} (1994), 39-54.

\bibitem{attouch} Attouch, H.: Variational Convergence for Functions and
Operators, \textit{Pitman Advanced Publishing Program}, Boston 1984.

\bibitem{AubFrank} Aubin, J.P. and Frankowska, H.: Set-Valued Analysis, 
\textit{Birkh\"{a}user}, Boston, 1990.

\bibitem{bak/polyak} Bakushinskii, A.B. and Poljak, B.T.: On the solution of
variational inequalities, \textit{Soviet Math. Dokl.} \textbf{15} (1974),
1705-1710.

\bibitem{BauBor-LegendreFcts} Bauschke, H.H. and Borwein, J.M.: Legendre
functions and the method of random Bregman projections. \textit{J. Convex
Anal.} \textbf{4} (1997) 27-67.

\bibitem{BauBorCom-Essential} Bauschke, H.H.; Borwein, J.M. and Combettes,
P.L.: Essential smoothness, essential strict convexity, and Legendre
functions in Banach spaces. \textit{Commun. Contemp. Math.} \textbf{3}
(2001), 615--647.

\bibitem{BauBorCom-BregMon} Bauschke, H.H., Borwein, J.M. and Combettes,
P.L.: Bregman monotone optimization algorithms. \textit{SIAM J. Control
Optim.} \textbf{42} (2003) 596-636.

\bibitem{BauCom-BregBestAp} Bauschke, H.H. and Combettes, P.L.: Construction
of best Bregman approximations in reflexive Banach spaces, \textit{Proc.
Amer. Math. Soc.} \textbf{131} (2003), 3757-3766.

\bibitem{BauComNoll-JointMin} Bauschke, H.H., Combettes, P.L. and Noll, D.:
Joint minimization with alternating Bregman proximity operators, preprint,
2005.

\bibitem{Bregman-Relaxation} Bregman, L.M.: The relaxation method for
finding common points of convex sets and its application to the solution of
problems in convex programming, \textit{USSR Computational Mathematics and
Mathematical Physics} \textbf{7} (1967), 200-217.

\bibitem{BonSha} Bonnans, J.F. and Shapiro, A.: Perturbation Analysis of
Optimization Problems, \textit{Springer Verlag}, New York, 2000.

\bibitem{BorZhu} Borwein, J.M. and Zhu, Q.J.: Techniques of Variational
Analysis, \textit{Springer}, New York 2005.

\bibitem{browder} Browder, F.E.: Multivalued monotone nonlinear mappings and
duality mappings in Banach spaces, \textit{Trans. Amer. Math. Soc. }\textbf{%
118 }(1965), 338-351.

\bibitem{browder1} Browder, F.E.: Existence and approximation of solutions
of nonlinear variational inequalities, \textit{Proc. Nat. Acad. Sci. U.S.A.} 
\textbf{56 (}1966), 1080--1086.

\bibitem{bro} Browder, F.E., Nonlinear Operators and Nonlinear Equations of
Evolution in Banach Spaces, "Proceedings of Symposia in Pure Mathematics",
Vol. XVIII, Part 2,\emph{\ }\textit{American Mathematical Society},
Providence, Rhode Island, 1976.

\bibitem{bruck1} Bruck, R.E.: An iterative solution for a variational
inequality for certain monotone operators in Hilbert space, \textit{Bull.
Amer. Math. Soc.} \textbf{81} (1975), 890-892. [Corrigendum: \textit{Bull.
Amer. Math. Soc.} \textbf{81 }(1976), p.353.]

\bibitem{BurSch} Burachik, R.S. and Scheimberg, S.: A proximal point method
for the variational inequality problem in Banach spaces, \textit{SIAM J.
Control Optim.} \textbf{39} (2000), 1633--1649.

\bibitem{ButIusProxPt} Butnariu, D. and Iusem, A.N.: On a proximal point
method for convex optimization in Banach spaces. \textit{Numer. Funct. Anal.
Optim.} \textbf{18} (1997), 723--744.

\bibitem{ButIus-Book} Butnariu, D. and Iusem, A.N.: Totally Convex Functions
for Fixed Points Computation and Infinite Dimensional Optimization, \textit{%
Kluwer Academic Publishers}, Dordrecht, 2000.

\bibitem{butnariu-ius-res} Butnariu, D., Iusem, A.N. and Resmerita, E.:
Total convexity for powers of the norm in uniformly convex Banach spaces, 
\textit{J. Convex Anal.} \textbf{7} (2000), 319-334.

\bibitem{ButIusZal} Butnariu, D., Iusem, A.N. and Z\u{a}linescu, C.: On
uniform convexity, total convexity and convergence of the proximal point and
outer Bregman projection algorithms in Banach spaces, \textit{J. Convex Anal.%
} \textbf{10} (2003), 1, 35--61.

\bibitem{ButResOptim} Butnariu, D. and Resmerita, E.: Averaged subgradient
methods for constrained convex optimization and Nash equilibria computation 
\textit{Optimization} \textbf{51} (2002), 863--888.

\bibitem{ButResSurvey} Butnariu, D. and Resmerita, E.: Bregman distances,
totally convex functions and a method for solving operator equations in
Banach spaces, \textit{Abstr. Appl. Anal.,} 2006, Art. ID 84919, 39 pages.

\bibitem{ButReiZas2003} Butnariu, D., Reich, S. and Zaslavski, A.: Weak
convergence for orbits of nonlinear operators in reflexive Banach spaces, 
\textit{Numer. Funct. Anal. Optim}. \textbf{24} (2003), 489-508.

\bibitem{czaszar} Csaszar, A.: General Topology, \textit{Akademiai Kiado},
Budapest, 1978.

\bibitem{DonZol} Dontchev, A.L. and Zolezzi, T.: Well-Posed Optimization
Problems, \textit{Springer Verlag}, Berlin, 1993.

\bibitem{eckstein} Eckstein, J.: Nonlinear proximal point algorithms using
Bregman functions, with application to convex programming, \textit{Math.
Operation Research} \textbf{18} (1993), 202-226.

\bibitem{eggermont} Eggermont, P.P.B.: Multiplicative iterative algorithms
for convex programming, \textit{Linear Algebra and Its Applications} \textbf{%
130 }(1990), 25-42.

\bibitem{eriksson} Eriksson, J.: An interval primal-dual algorithm for
linear programming, \textit{Technical Report 85-10}, \textit{Department of
Mathematics, Link\"{o}ping University}, Sweden, 1985.

\bibitem{erlander} Erlander, S.: Entropy in linear programs, \textit{Math.
Programming} \textbf{21 }(1981), 137-151.

\bibitem{ermoliev} Ermoliev, Yu.M.: Methods for solving nonlinear extremal
problems, \textit{Kibernetika (Kiev)} \textbf{1} (1966), 1-17 (Russian).

\bibitem{FacPan} Facchinei, F. and Pang, J.-S.: Finite-Dimensional
Variational Inequalities and Complementarity Problems, Vol. 1, \textit{%
Springer Verlag}, New York, 2003.

\bibitem{GoeRei} Goebel, K. and Reich, S.: Uniform Convexity, Hyperbolic
Geometry, and Nonexpansive Mappings, \textit{Marcel Dekker}, New York and
Basel, 1984.

\bibitem{kassay} Kassay, G.: The proximal point algorithm for reflexive
Banach spaces, \textit{Studia Math.} \textbf{30} (1985), 9-17.

\bibitem{krasnoselskii} Krasnoselskii, M.A.: Two observations about the
method of succesive approximations (Russian), \textit{Uspekhi
Mathematicheskikh Nauk} \textbf{10} (1955), 123-127.

\bibitem{landweber} Landweber, L.: An iterative formula for Fredholm
integral equations of the first kind, \textit{Amer. J. Math.} \textbf{73}
(1951), 615-624.

\bibitem{lemaire} Lemaire, B.: The proximal algorithm, in: J.P. Penot (Ed.),
"International Series of Numerical Mathematics", No. 87, pp. 83-97, \textit{%
Birkh\"{a}user}, Basel, 1989.

\bibitem{liskovets3} Liskovets, O. A.: External approximations for the
regularization of monotone variational inequalities, \textit{Soviet Math.
Dokl.} \textbf{36} (1988), 220-224.

\bibitem{liskovets4} Liskovets, O. A.: Regularized variational inequalities
with pseudo-monotone operators on approximately given sets (Russian). 
\textit{Differential Equations} \textbf{11} (1989), 1970-1977.

\bibitem{martinet} Martinet, B.: R\'{e}gularization d'in\'{e}quations
variationelles par approximations successive, \textit{Revue Fran\c{c}aise de
Informatique et Recherche Op\'{e}rationelle} \textbf{2 (}1970), 154-159.

\bibitem{martinet-1} Martinet, B.: Algorithmes pour la r\'{e}solution des
probl\`{e}mes d'optimisation et minimax, Th\`{e}se d'\'{e}tat, \textit{%
Universit\'{e} de Grenoble}, Grenoble, France, 1972.

\bibitem{Moreau1962} Moreau, J.-J.: Fonctions convexes duales et points
proximaux dans un espace hilbertien\textit{.} \textit{C. R. Acad. Sc. Paris} 
\textbf{255} (1962), 2897-2899.

\bibitem{Moreau1963} Moreau, J.-J.: Propri\'{e}t\'{e}s des applications
`prox', \textit{C. R. Acad. Sc. Paris} \textbf{256} (1963), 1069-1071.

\bibitem{Moreau1965} Moreau, J.-J.: Proximit\'{e} et dualit\'{e} dans un
espace hilbertien, \textit{Bull. Soc.} \textit{Math. France} \textbf{93 (}%
1965), 273-299.

\bibitem{mosco} Mosco, U.: Convergence of convex sets and of solutions of
variational inequalities, \textit{Advances in Math.} \textbf{3} (1969),
510-585.

\bibitem{opial} Opial, Z.: Weak convergence of the sequence of successive
approximations for nonexpansive mappings, \textit{Bull. Amer. Math. Soc.} 
\textbf{73} (1967), 591-597.

\bibitem{PasSbu} Pascali, D. and Sburlan, S.: Nonlinear Mapping of Monotone
Type, \textit{Martinus Nijhoff}, Dordrecht, 1978.

\bibitem{PhelpsBook} Phelps, R.R.: Convex Functions, Monotone Operators, and
Differentiability - 2nd Edition. \textit{Springer Verlag}, Berlin, 1993.

\bibitem{polyak} Polyak, B.T.: A general method for solving extremum
problems, \textit{Dokl.} \textit{Akad. Nauk SSSR} \textbf{174} (1967),
593-597.

\bibitem{polyakBook} Polyak, B.T.: Introduction to Optimization, \textit{%
Optimization Software, Inc., Publications Division,} New York, 1987.

\bibitem{RocLevelSets} Rockafellar, R.T.: Level sets and continuity of
conjugate convex functions. \textit{Trans. Amer. Math. Soc.} \textbf{123 }%
(1966), 46-63.

\bibitem{RocSums} Rockafellar, R.T.: On the maximality of sums of nonlinear
operators. \textit{Trans. Amer. Math. Soc.} \textbf{49} (1970), 75-88.

\bibitem{RocProxPt} Rockafellar, R.T.: Monotone operators and the proximal
point algorithm. \textit{SIAM J. Control Optim.} \textbf{14} (1976),
877--898.

\bibitem{RocAugLag} Rockafellar, R.T.: Augmented Lagrangians and
applications of the proximal point algorithm in convex programming. \textit{%
Math. Oper. Res.} \textbf{1} (1976), 97--116.

\bibitem{RocWetts-Book} Rockafellar, R.T. and Wetts, R.J.-B.: Variational
Analysis. \textit{Springer Verlag}, Berlin, 1998.

\bibitem{ruszczynski} Ruszczy\'{n}ski, A.: A merit function approach of the
subgradient method, \textit{preprint}, 2006.

\bibitem{shor1962} Shor, N.Z.: Application of the method of gradient descent
to the solution of the network transportation problem, in: "Materials of the
Scientific Seminar on Theoretical and Applied Questions of Cybernetics and
Operation Research", \textit{Ukrainian Academy of Sciences}, Kiev, 1962, pp.
1-17 (Russian).

\bibitem{shorbook} Shor, N.Z.: Minimization Methods for Non-Differentiable
Functions, \textit{Springer Verlag}, 1985.

\bibitem{tikhonov} Tikhonov, A.N.: Regularization of incorrectly posed
problems, \textit{Soviet Math. Dokl.} \textbf{4} (1963), 1035-1038.

\bibitem{yoshida} Yosida, K.: Lectures of Differential and Integral
Equations, \textit{Interscience,} London, 1960.
\end{thebibliography}
\end{document}